# Bayesian Checking of the Second Levels of Hierarchical Models[1]

**M. J. Bayarri and M. E. Castellanos**

*Abstract.* Hierarchical models are increasingly used in many applications. Along with this increased use comes a desire to investigate whether the model is compatible with the observed data. Bayesian methods are well suited to eliminate the many (nuisance) parameters in these complicated models; in this paper we investigate Bayesian methods for model checking. Since we contemplate model checking as a preliminary, exploratory analysis, we concentrate on *objective Bayesian methods* in which careful specification of an informative prior distribution is avoided. Numerous examples are given and different proposals are investigated and critically compared.

## 1. INTRODUCTION

With the availability of powerful numerical computations, use of hierarchical (or multilevel, or random effects) models has become very common in applications. They nicely generalize and extend standard one-level models to complicated situations, where these simple models would not apply. With their widespread use comes along an increased need to check the adequacy of such models to the observed data. Recent Bayesian methods (Bayarri and Berger (1999), 2000) have shown con-siderable promise in checking one-level models, especially in nonstandard situations in which parameter-free testing statistics are not known. In this paper we show how these methods can be extended to checking hierarchical models. We also review state-of-the-art Bayesian proposals for checking hierarchical models and critically compare them.

*M. J. Bayarri is Professor, Department of Statistics and Operation Research, University of Valencia, Burjassot, (Valencia), 46100 Spain e-mail: susie.bayarri@uv.es. M. E. Castellanos is Associate Professor, Department of Statistics and Operation Research, Rey Juan Carlos University, Móstoles, (Madrid), 28933 Spain e-mail: maria.castellanos@urjc.es.*

We approach model checking as a preliminary analysis in that if the data are compatible with the assumed model, then the full (and difficult) Bayesian process of model elaboration and model selection (or averaging) can be avoided. The role of Bayesian model checking versus model selection has been discussed, for example, in Bayarri and Berger (1999, 2000) and O'Hagan (2003) and we will not repeat it here.

In general, in a parametric model checking scenario, we relate observables $\mathbf{X}$ with parameters $\boldsymbol{\theta}$ through a parametric model $\mathbf{X} \mid \boldsymbol{\theta} \sim f(\mathbf{x} \mid \boldsymbol{\theta})$. We then observe data $\mathbf{x}_{obs}$ and wish to assess whether $\mathbf{x}_{obs}$ are compatible with the assumed (null) model $f(\mathbf{x} \mid \boldsymbol{\theta})$. Most of the existing methods for model checking (both Bayesian and frequentist) can be seen to correspond to particular choices of:

1. A diagnostic statistic $T$, to quantify incompatibility of the model with the observed data through $t_{obs} = T(\mathbf{x}_{obs})$.

2. A completely specified distribution for the statistic, $h(t)$, under the null model, in which to locate the observed $t_{obs}$.

3. A way to measure conflict between the observed statistic, and the null distribution, $h(t)$, for $T$. The most popular measures are tail areas ($p$-values) and relative height of the density $h(t)$ at $t_{obs}$.

In this paper, we concentrate on the optimal choice in item 2, which basically reduces to choice of methods to eliminate the nuisance parameters $\boldsymbol{\theta}$ from the null model. Our recommendations then apply to *any* choices in 1 and 3. [We abuse notation and use the same $h(\cdot)$ to indicate both the completely specified distribution for $\mathbf{X}$, after elimination of $\boldsymbol{\theta}$, and the corresponding distribution for $T$.] Of course, choice of 1 is very important; as a matter of fact, in some scenarios a "good" $T$ can be chosen which is ancillary or nearly so, thus making choice of 2 nearly irrelevant. So our work will be most relevant for complicated scenarios when such optimal $T$'s are not known, or extremely difficult to implement (for an example of these, see Robins, van der Vaart and Ventura (2000)). In these situations, $T$ is often chosen casually, based on intuitive considerations, and hence we concentrate on these choices (with *no* implications whatsoever that these are our recommended choices for $T$; we simply do not address choice of $T$ in this paper). Also, without loss of generality, we can assume that $T$ has been defined such that the larger $T$ is, the more incompatible data are with the assumed model. As measures of conflict in item 3 above, we explore the two best known *measures of surprise*, namely the $p$-value and the *relative predictive surprise, RPS* (see Berger (1985), Section 4.7.2) used (with variants) by many authors. These two measures are defined as

$$(1.1) \qquad p = Pr^{h(\cdot)}(t(\mathbf{X}) \geq t(\mathbf{x}_{obs})),$$

$$(1.2) \qquad RPS = \frac{h(t(\mathbf{x}_{obs}))}{\sup_t h(t)}.$$

Note that *small* values of (1.1) and (1.2) denote incompatibility.

Frequentist and Bayesian choices for $h(\cdot)$ are discussed at length in Bayarri and Berger (2000), and we limit ourselves here to an extremely brief (and incomplete) mention of some of them. The natural Bayesian choice for $h(\cdot)$ is the prior predictive distribution, in which the parameters get naturally integrated out with respect to the prior distribution. (Box (1980) pioneered use of $p$-values computed in the prior predictive for Bayesian model criticism.) However, this requires a fairly informative prior distribution (see O'Hagan (2003) for a discussion) which might not be desirable for model checking for two reasons: (i) we might wish to avoid the careful (and difficult) prior quantification in these earlier stages of the analysis, since the model might well not be appropriate and hence the effort is wasted; (ii) most importantly, model checking with informative priors cannot separate inadequacy of the prior from inadequacy of the model.

In the sequel we will concentrate on *objective* Bayesian methods for model checking. We use the term *objective* to refer to Bayesian methods in which the priors are chosen by some default, agreed upon rules (*objective priors*) rather than reflecting genuine (subjective) prior information. This term is frequent among Bayesians (see, e.g., Berger, 2003, 2006) but its use is not without controversy. Objective priors are usually improper. Note that this impropriety makes the prior predictive distribution undefined and hence not available for (objective) model checking.

Guttman's (1967) and Rubin's (1984) choice for $h(\cdot)$ is the *posterior* predictive distribution, resulting from integrating $\boldsymbol{\theta}$ out with respect to the posterior distribution instead of the prior. This allows use of improper priors, and hence of objective model checking. This proposal is very easy to implement by Markov chain Monte Carlo (MCMC) methods, and hence has become fairly popular in Bayesian model checking. However, its double use of the data can result in an extreme conservatism of the resulting $p$-values, unless the checking statistic is fairly ancillary (in which case the way to deal with the parameters is basically irrelevant). This conservatism is shown to hold asymptotically in Robins, van der Vaart and Ventura (2000), and for finite $n$ and several scenarios in, for example, Bayarri and Berger (1999, 2000), Bayarri and Castellanos (2001) and Bayarri and Morales (2003). Miscalibration of posterior predictive measures is also documented in Dahl (2006), Draper and Krnjajić (2006) and Hjort, Dahl and Steinbakk (2006); the double use of the data was noted in the discussion of Gelman, Meng and Stern (2003) (see, in particular, Draper (1996)). This is not meant in any way to imply that posterior predictive measures are without merit [see Gelman (2003) for a recent exposition of their advantages and interpretation], only that they have to be interpreted in a different way: a posterior $p$-value equal to, say, 0.4 can *not* naively be interpreted as compatibility with the null model in *all* problems. A small posterior predictive measure can safely be interpreted as incompatibility with the null model.

Alternative choices of $h(\cdot)$ for objective model checking are proposed in Bayarri and Berger (1997, 1999, 2000). Their asymptotic optimality is shown in Robins, van der Vaart and Ventura (2000). In this paper we derive these marginals for hierarchical model checking. We also compare the results with those obtained with posterior predictive distributions and several "plug-in" choices for $h(\cdot)$. Note that "plug-in" $p$-values would be natural choices for frequentist checking when interpreting the second level of a hierarchical model as a "random effect," so in particular, we compare some popular choices of Bayesian $p$-values with MLE "plug-in" $p$-values.

There are not many proposals for checking the distributional assumption of "random effects." Along with the mentioned methods, we also carefully review state-of-the-art Bayesian proposals, namely (i) the *simulation-based* checking of Dey, Gelfand, Swartz and Vlachos (1998), a computationally intensive method based on Monte Carlo tests, (ii) the *O'Hagan* method (O'Hagan (2003)) for checking graphical models, and (iii) the *conflict $p$-values* of Marshall and Spiegelhalter (2003), close in spirit to cross-validation methods. We critically compare these methods in several examples. In this paper most attention is devoted to the checking of a fairly simple normal-normal hierarchical model so as to best illustrate the different proposals and critically judge their behavior. Of course, the main ideas also apply to the checking of many other hierarchical models. In Section 2 we briefly review the different *measures of surprise* (MS) that we will derive and compare. In Section 3 we derive these measures for the hierarchical normal-normal model. We also study the sampling distribution of the different $p$-values, both when the null model is true, and when the data come from alternative models. In Section 4 we apply these measures to a particular simple test which allows easy and intuitive comparisons of the different proposals. In Section 5 we briefly summarize other methods for Bayesian checking of hierarchical models, namely those proposed by Dey, Gelfand, Swartz and Vlachos (1998), O'Hagan

(2003) and Marshall and Spiegelhalter (2003), comparing them with the previous proposals in an example. Finally, in Section 6 we check the adequacy of a binomial/beta hierarchical model in a well-known example using all of the methods reviewed in the paper.

## 2. MEASURES OF SURPRISE IN THE CHECKING OF HIERARCHICAL MODELS

In this paper we will be dealing with the MS defined in (1.1) and (1.2). Their relative merits and drawbacks are discussed at length in Bayarri and Berger (1997, 1999) and will not be repeated here. In this section we derive these measures in the context of hierarchical models, and for some specific choices of the completely specified distribution $h(\cdot)$. We consider the general two-level model:

$$X_{ij} \mid \theta_i \stackrel{ind.}{\sim} f(x_{ij} \mid \theta_i), \quad i = 1, \ldots, I; j = 1, \ldots, n_i,$$

$$\boldsymbol{\theta} \mid \boldsymbol{\eta} \stackrel{ind.}{\sim} \pi(\boldsymbol{\theta} \mid \boldsymbol{\eta}) = \prod_{i=1}^{I} \pi(\theta_i \mid \boldsymbol{\eta}),$$

$$\boldsymbol{\eta} \sim \pi(\boldsymbol{\eta}),$$

where $\boldsymbol{\theta} = (\theta_1, \ldots, \theta_I)$ and $\boldsymbol{\eta} = (\eta_1, \ldots, \eta_p)$

To get a completely specified distribution $h(\cdot)$ for $\mathbf{X}$, we need to integrate $\boldsymbol{\theta}$ out from $f(\mathbf{x} \mid \boldsymbol{\theta})$ with respect to some completely specified distribution for $\boldsymbol{\theta}$. We next consider three possibilities that have been proposed in the literature for such a distribution: empirical Bayes types (plug-in), posterior distribution, and partial posterior distribution, as they apply in the hierarchical scenario. Notice that, since we will be dealing with improper priors for $\boldsymbol{\eta}$, the natural (marginal) prior $\pi(\boldsymbol{\theta})$ is also improper and cannot be used for this purpose [it would produce an improper $h(\cdot)$].

### 2.1 Empirical Bayes (Plug-In) Measures

This is the simplest proposal, very intuitive and frequently used in empirical Bayes analysis (see, e.g., Carlin and Louis (2000), Chapter 3). It simply consists in replacing the unknown $\boldsymbol{\eta}$ in $\pi(\boldsymbol{\theta} \mid \boldsymbol{\eta})$ by an estimate (we use the MLE, but moment estimates are often used as well). In this proposal, $\boldsymbol{\theta}$ is integrated out with respect to

$$(2.1) \qquad \pi^{EB}(\boldsymbol{\theta}) = \pi(\boldsymbol{\theta} \mid \hat{\boldsymbol{\eta}}) = \pi(\boldsymbol{\theta} \mid \boldsymbol{\eta} = \hat{\boldsymbol{\eta}}),$$

where $\hat{\boldsymbol{\eta}}$ maximizes the integrated likelihood:

$$f(\mathbf{x} \mid \boldsymbol{\eta}) = \int f(\mathbf{x} \mid \boldsymbol{\theta}) \pi(\boldsymbol{\theta} \mid \boldsymbol{\eta}) \, d\boldsymbol{\theta}.$$

The corresponding proposal for a completely specified $h(\cdot)$ in which to define the MS is

$$(2.2) \qquad m_{prior}^{EB}(t) = \int f(t \mid \boldsymbol{\theta}) \pi^{EB}(\boldsymbol{\theta}) \, d\boldsymbol{\theta}.$$

The MS $p_{prior}^{EB}$ and $RPS_{prior}^{EB}$ are now given by (1.1) and (1.2), respectively, in which $h(\cdot) = m_{prior}^{EB}(\cdot)$.

Strictly for comparison purposes, we will later use another distribution which is also of the empirical Bayes type; in this new distribution, the empirical Bayes prior (2.1) gets needlessly (and inappropriately) updated using again the same data. In this (wrong) proposal, $\boldsymbol{\theta}$ gets integrated out with respect to

$$(2.3) \qquad \pi^{EB}(\boldsymbol{\theta} \mid \mathbf{x}_{obs}) \propto f(\mathbf{x}_{obs} \mid \boldsymbol{\theta}) \pi^{EB}(\boldsymbol{\theta}),$$

resulting in

$$(2.4) \quad m_{post}^{EB}(t) = \int f(t \mid \boldsymbol{\theta}) \pi^{EB}(\boldsymbol{\theta} \mid \mathbf{x}_{obs}) \, d\boldsymbol{\theta}.$$

The corresponding MS $p_{post}^{EB}$ and $RPS_{post}^{EB}$ are again computed using (1.1) and (1.2), respectively, with $h(\cdot) = m_{post}^{EB}(t)$.

## 2.2 Posterior Predictive Measures

This proposal is also intuitive and seems to have a more Bayesian "flavour" than the plug-in solution presented in the previous section. This along with its ease of implementation has made the method a popular one for objective Bayesian model checking. This popularity makes investigation of its properties all the more important. The idea is simple: use the posterior to integrate $\boldsymbol{\theta}$ out. Assuming that the posterior is proper (as usual), this allows model checking when $\pi(\boldsymbol{\eta})$ [and hence $\pi(\boldsymbol{\theta})$] is improper. Thus, the proposal for $h(\cdot)$ is the posterior predictive distribution

$$(2.5) \ m_{post}(t \mid \mathbf{x}_{obs}) = \int f(t \mid \boldsymbol{\theta}) \pi(\boldsymbol{\theta} \mid \mathbf{x}_{obs}) \, d\boldsymbol{\theta},$$

where $\pi(\boldsymbol{\theta} \mid \mathbf{x}_{obs})$ is the marginal from the joint posterior

$$\pi(\boldsymbol{\theta}, \boldsymbol{\eta} \mid \mathbf{x}_{obs}) \propto f(\mathbf{x}_{obs} \mid \boldsymbol{\theta}) \pi(\boldsymbol{\theta}, \boldsymbol{\eta})$$

$$= f(\mathbf{x}_{obs} \mid \boldsymbol{\theta}) \pi(\boldsymbol{\eta}) \prod_{i=1}^{I} \pi(\theta_i \mid \boldsymbol{\eta}).$$

The *posterior p-value* and the posterior *RPS* are denoted by $p_{post}$ and $RPS_{post}$, and computed from (1.1) and (1.2), respectively, with $h(\cdot) = m_{post}(\cdot)$.

It is important to remark that, under regularity conditions, the empirical Bayes posterior $\pi^{EB}(\boldsymbol{\theta} \mid \mathbf{x}_{obs})$ given in (2.3) approximates the true posterior $\pi(\boldsymbol{\theta} \mid \mathbf{x}_{obs})$. Both are, in fact, asymptotically equivalent. Hence whatever inadequacy of $m_{post}^{EB}(t)$ in (2.4) for model checking is likely to apply as well to the posterior predictive $m_{post}(t \mid \mathbf{x}_{obs})$ in (2.5). We will see demonstration of the similar behavior of both predictive distributions in all the examples in this paper. Use of posterior predictive measures was introduced by Guttman (1967) and Rubin (1984) and extended and formalized in Gelman, Meng and Stern (2003). They are very easy to compute and they are perhaps the most widely used checking procedure. We refer to Meng (1994), Gelman, Meng and Stern (2003) and Gelman (2003) for extended discussion and motivation.

## 2.3 Partial Posterior Predictive Measures

Both the empirical Bayes and posterior proposals presented in Sections 2.1 and 2.2 use the *same* data to (i) "train" the improper $\pi_0^{EB}$ into a proper distribution to compute a predictive distribution and (ii) compute the observed $t_{obs}$ to be located in this *same* predictive distribution through the MS. This can result in a severe conservatism incapable of detecting clearly inappropriate models. A natural way to avoid this double use of the data is to use part of the data for "training" and the rest to compute the MS, as in cross-validation methods. The proposal in Bayarri and Berger (1999, 2000) is similar in spirit: since $t_{obs}$ is used to compute the surprise measures, it uses the information in the data *not* in $t_{obs}$ to "train" the improper prior into a proper one. A natural way to "remove" the information in $t_{obs} = T(\mathbf{X} = \mathbf{x}_{obs})$ from $\mathbf{x}_{obs}$ is by conditioning in the observed value of the statistic $T(\mathbf{X})$; that is, using the conditional distribution $f(\mathbf{x}_{obs} \mid t_{obs}, \boldsymbol{\theta})$ instead of $f(\mathbf{x}_{obs} \mid \boldsymbol{\theta})$ to define the likelihood. The resulting posterior distribution for $\boldsymbol{\theta}$ (assumed proper) is called a *partial posterior distribution* and given by

$$\pi_{ppp}(\boldsymbol{\theta} \mid \mathbf{x}_{obs} \setminus t_{obs}) \propto f(\mathbf{x}_{obs} \mid t_{obs}, \boldsymbol{\theta}) \pi(\boldsymbol{\theta})$$

$$\propto \frac{f(\mathbf{x}_{obs} \mid \boldsymbol{\theta}) \pi(\boldsymbol{\theta})}{f(t_{obs} \mid \boldsymbol{\theta})}.$$

The corresponding proposal for the completely specified $h(\cdot)$ is then the *partial posterior predictive distribution* computed as

$$m_{ppp}(t \mid \mathbf{x}_{obs} \setminus t_{obs}) = \int f(t \mid \boldsymbol{\theta}) \pi(\boldsymbol{\theta} \mid \mathbf{x}_{obs} \setminus t_{obs}) \, d\boldsymbol{\theta}.$$

The *partial posterior predictive measures* of surprise will be denoted by $p_{ppp}$ and $RPS_{ppp}$ and, as before,

computed using (1.1) and (1.2), respectively, with $h(\cdot) = m_{ppp}(\cdot)$.

Extensive discussions of the advantages and disadvantages of this proposal as compared with the previous ones can be found in Bayarri and Berger (2000) and Robins, van der Vaart and Ventura (2000). In this paper we demonstrate their performance in hierarchical models.

## 2.4 Computation of $p_{h(\cdot)}$ and $RPS_{h(\cdot)}$

Often, for a proposed $h(\cdot)$, the measures $p_{h(\cdot)}$ and $RPS_{h(\cdot)}$ cannot be computed in closed form. In fact, $h(\cdot)$ is often not of closed form itself. In these cases we use Monte Carlo (MC), or Markov Chain Monte Carlo (MCMC) methods, to (approximately) compute them. If $\mathbf{x}^1, \ldots, \mathbf{x}^M$ is a sample of size $M$ generated from $h(\mathbf{x})$, then $t_i = t(\mathbf{x}^i)$ is a sample from $h(t)$, and we approximate the MS as:

1. $p$-value

$$Pr^{h(\cdot)}(T \geq t_{obs}) = \frac{\# \text{ of } t_i \geq t_{obs}}{M},$$

2. relative predictive surprise

$$RPS_{h(\cdot)} = \frac{\hat{h}(t_{obs})}{\sup_t \hat{h}(t)},$$

where $\hat{h}(t)$ is an estimate (for instance a kernel estimate) of the density $h$ at $t$. When the distribution of the test statistic $T$, $f_T(t \mid \boldsymbol{\theta})$, has closed form expression, one can avoid kernel estimation by using a "Rao–Blackwellized" Monte Carlo estimate of $h$, that is, $\hat{h}(t) = (1/m) \sum_{k=1}^{m} f_T(t \mid \boldsymbol{\theta}_k)$, where the $\boldsymbol{\theta}_k$'s are draws from the appropriate distribution for $\boldsymbol{\theta}$ (proper prior, posterior, partial posterior, …). This is the method used in the examples of this paper and was pointed to us by a referee.

## 3. CHECKING HIERARCHICAL NORMAL MODELS

Consider the usual normal-normal two-level hierarchical (or random effects) model with $I$ groups and $n_i$ observations per group. The $I$ means are assumed to be exchangeable. For simplicity, we begin by assuming the variances $\sigma_i^2$ at the observation level to be known. The model is

$$X_{ij} \mid \theta_i \overset{i}{\sim} N(\theta_i, \sigma_i^2),$$

$$(3.1) \qquad i = 1, \ldots, I, j = 1, \ldots, n_i,$$

$$\pi(\boldsymbol{\theta} \mid \mu, \tau) = \prod_{i=1}^{I} N(\theta_i \mid \mu, \tau^2),$$

$$\pi(\mu, \tau^2) = \pi(\mu)\pi(\tau^2) \propto \frac{1}{\tau}.$$

In this paper we concentrate on checking the adequacy of the second-level assumptions on the means $\theta_i$. Of course, checking the normality of the observations is also important, but it will not be considered here. The techniques considered in this paper as applied to the checking of simple models have been explored in Bayarri and Castellanos (2001), Castellanos (1999) and Bayarri and Morales (2003).

Assume that choice of the departure statistic $T$ is done in a rather casual manner, and that we are especially concerned about the upper tail of the distribution of the means. In this situation, a natural choice for $T$ is

$$(3.2) \qquad T = \max\{\overline{X}_1., \ldots, \overline{X}_I.\},$$

where $\overline{X}_i.$ denotes the usual sample mean for group $i$. This $T$ is rather natural, but the analysis would be virtually identical with any other choice. Recall that if the statistic is fairly ancillary, then the answers from all methods are going to be rather similar, no matter how we integrate $\boldsymbol{\theta}$ out.

The density of the statistic (3.2) under the (null) model specified in (3.1) can be computed to be

$$
\begin{aligned}
(3.3) \quad f_T(t \mid \boldsymbol{\theta}) &= \sum_{k=1}^{I} N\left(t \mid \theta_k, \frac{\sigma_k^2}{n_k}\right) \\
&\quad \cdot \prod_{\substack{l=1 \\ l \neq k}}^{I} F\left(t \mid \theta_l, \frac{\sigma_l^2}{n_l}\right),
\end{aligned}
$$

where $N(t \mid a, b)$ and $F(t \mid a, b)$ denote the density and distribution function, respectively, of a normal distribution with mean $a$ and variance $b$ evaluated at $t$.

We next integrate the unknown $\boldsymbol{\theta}$ from (3.3) using the techniques outlined in Section 2.

### 3.1 Empirical Bayes Distributions

It is easy to see that the likelihood for $\mu$ and $\tau^2$ is simply

$$(3.4) \qquad f(\mathbf{x} \mid \mu, \tau^2) = \prod_{i=1}^{I} N\left(\bar{x}_i \mid \mu, \frac{\sigma_i^2}{n_i} + \tau^2\right),$$

from which $\hat{\mu}$ and $\hat{\tau}^2$ can be computed. Then (2.1) is given by

$$\pi^{EB}(\boldsymbol{\theta}) = \pi(\boldsymbol{\theta} \mid \hat{\mu}, \hat{\tau}^2) = \prod_{i=1}^{I} N(\theta_i \mid \hat{\mu}, \hat{\tau}^2),$$

which we use to integrate $\boldsymbol{\theta}$ out from (3.3). The resulting $m_{prior}^{EB}(\mathbf{x})$ does not have a closed form expression, but simulations can be obtained by simple MC methods. For comparison purposes, we will also consider integrating $\boldsymbol{\theta}$ w.r.t. the (inappropriate) empirical Bayes posterior distribution. The resulting $m_{post}^{EB}(\mathbf{x})$ is also trivial to simulate from by using a similar MC scheme. Details are given in Appendix A.

### 3.2 Posterior Predictive Distribution

This proposal integrates $\boldsymbol{\theta}$ out from (3.3) w.r.t. its posterior distribution. For the noninformative prior $\pi(\mu, \tau^2) \propto 1/\tau$, the joint posterior is

$$\begin{aligned}
&\pi_{post}(\boldsymbol{\theta}, \mu, \tau^2 | \mathbf{x}_{obs}) \\
(3.5) \quad &\propto f(\mathbf{x} \mid \boldsymbol{\theta}, \mu, \tau^2) \pi(\boldsymbol{\theta} \mid \mu, \tau^2) \pi(\mu, \tau^2) \\
&= \frac{1}{\tau} \prod_{i=1}^{I} N\left(\overline{x}_{i\cdot} \mid \theta_i, \frac{\sigma_i^2}{n_i}\right) \prod_{i=1}^{I} N(\theta_i \mid \mu, \tau^2).
\end{aligned}$$

To simulate from the resulting posterior predictive distribution $m_{post}(\mathbf{x} \mid \mathbf{x}_{obs})$, we first simulate from $\pi_{post}(\boldsymbol{\theta}, \mu, \tau^2 | \mathbf{x}_{obs})$ and for each simulated $\boldsymbol{\theta}$, we simulate $\mathbf{x}$ from $f(\mathbf{x} \mid \boldsymbol{\theta})$. To simulate from the joint posterior (3.5) we use an easy Gibbs sampler defined by full conditionals given in Appendix B.

### 3.3 Partial Posterior Distribution

To simulate from the partial posterior predictive distribution, $m_{ppp}$, we proceed similarly to Section 3.2, except that simulations for the parameters are generated from the partial posterior distribution:

$$\pi_{ppp}(\boldsymbol{\theta}, \mu, \tau^2 \mid \mathbf{x}_{obs} \setminus t_{obs}) \propto \frac{\pi_{post}(\boldsymbol{\theta}, \mu, \tau^2 \mid \mathbf{x}_{obs})}{f(t_{obs} \mid \boldsymbol{\theta})},$$

where $\pi_{post}(\boldsymbol{\theta}, \mu, \tau^2 \mid \mathbf{x}_{obs})$ is given in (3.5). Details are given in Appendix C.

### 3.4 Examples

For illustration, we now compute the MS, that is, the $p$-values and the relative predictive surprise indexes for the different proposals. We use a couple of data sets with five groups and eight observations in each group. In both of them the null model is not the model generating the data; in Example 1 one of the means comes from a different normal with a larger mean, whereas in Example 2 the means come from a Gamma distribution. Recall that the null model (3.1) had the group means i.i.d. normal.

EXAMPLE 1.  The group means are 1.56, 0.64, 1.98, 0.01, 6.96, simulated from

$$\begin{aligned}
&X_{ij} \sim N(\theta_i, 4), \quad i = 1, \ldots, 5, j = 1, \ldots, 8, \\
&\theta_i \sim N(1, 1), \quad j = 1, \ldots, 4, \\
&\theta_5 \sim N(5, 1).
\end{aligned}$$

EXAMPLE 2.  The group means are: 0.75, 0.77, 5.77, 1.86, 0.75, simulated from

$$\begin{aligned}
&X_{ij} \sim N(\theta_i, 4), \quad i = 1, \ldots, 5, j = 1, \ldots, 8, \\
&\theta_i \sim Ga(0.6, 0.2), \quad i = 1, \ldots, 5.
\end{aligned}$$

In Table 1 we show all MS for the two examples. The partial posterior measures clearly detect the inadequate models, with very small $p$-values and $RPS$. On the other hand, none of the other predictive distributions work well for this purpose, no matter how we choose to locate the observed $t_{obs}$ in them (with $p$-values or $RPS$). The prior empirical Bayes are conservative, with $p$ and $RPS$ an order of magnitude larger than the ones produced by the partial posterior predictive distribution. Both the posterior empirical Bayes and predictive posterior measures are *extremely* conservative, indicating almost perfect agreement of the observed data with the quite obviously wrong null models. Besides, it can be seen that EB posterior and posterior predictive measures are very similar to each other. This is not a specific feature of these examples, but occurs very often. We further explore it in a rather simple null model in Section 4.

We next study the behavior of the different $p$-values, when considered as a function of $\mathbf{X}$, under the null and under some alternatives.

### 3.5 Null Sampling Distribution of the $p$-Values

In Section 2, we have reviewed four different ways to define (Bayesian) $p$-values for model checking. To compare their performance, we should address the question of what do we want in a $p$-value.

For frequentists, one appealing property of $p$-values is that, when considered as random variables, $p(\mathbf{X})$ have $U(0, 1)$ distributions under the null models. This endorses $p$-values with a very desirable property, namely having the same interpretation across problems. Statistical measures that lack a common interpretation across problems are simply not very useful. (For more extensive discussion of this point, see Robins, van der Vaart and Ventura (2000).) In fact, the uniformity of *p*-values has often been taken as their "defining" characteristic (Meng (1994); Rubin (1996); De la Horra and Rodríguez-Bernal (1997); Thompson (1997); Robins (1999); Robins, van der Vaart and Ventura (2000)). For most problems, exact uniformity under the null for all $\boldsymbol{\theta}$ cannot be attained for any *p*-value. Thus one must weaken the requirement to some extent. A natural weaker requirement is that a *p*-value be $U(0, 1)$ under the null in an asymptotic sense (see Robins, van der Vaart and Ventura (2000)). As an aside, it should be remarked that uniformity of *p*-values is an essential assumption for some analyses based on *p*-values, as some popular algorithms for handling multiplicities (see Cabras (2004)).

TABLE 1
*p-values and RPS for Examples 1 and 2*

|        | $p_{prior}^{EB}$ | $RPS_{prior}^{EB}$ | $p_{post}^{EB}$ | $RPS_{post}^{EB}$ | $p_{post}$ | $RPS_{post}$ | $p_{ppp}$ | $RPS_{ppp}$ |
|--------|--------|--------|--------|--------|--------|--------|--------|--------|
| Ex. 1  | 0.13   | 0.28   | 0.35   | 0.93   | 0.41   | 0.97   | 0.01   | 0.01   |
| Ex. 2  | 0.12   | 0.29   | 0.30   | 0.88   | 0.38   | 0.95   | 0.01   | 0.01   |

It is not obvious that Bayesians should be concerned with establishing that a *p*-value is uniform under the null for all $\boldsymbol{\theta}$. For instance, when the prior is proper, the prior predictive *p*-value is $U(0, 1)$ under $m(\mathbf{x})$, which means it is $U(0, 1)$ in an average sense over $\boldsymbol{\theta}$. If the prior distribution is chosen subjectively, a Bayesian could well argue that this is sufficient. Indeed Meng (1994) suggested that uniformity under $m(\mathbf{x})$ is a useful criterion for the evaluation of any proposed (Bayesian) *p*-value.

If the prior is improper, however (as it is often the case in objective Bayes model checking, the subject of this paper), then this prior predictive uniformity criterion cannot be used. Of course, if a *p*-value is uniform under the null in the frequentist sense, then it has the strong Bayesian property of being marginally $U(0, 1)$ under *any* proper prior distribution. This explains why Bayesians should, at least, be highly satisfied if the frequentist requirement obtains. Perhaps more to the point, if a proposed *p*-value is *always* either conservative or anticonservative in a frequentist sense (see Robins, van der Vaart and Ventura (2000), for definitions), then it is likewise guaranteed to be conservative or anti-conservative in a Bayesian sense, no matter what the prior. Interesting related discussion concerning the posterior predictive *p*-value can be found in Meng (1994), Gelman, Meng and Stern (2003), Rubin (1996), Gelman (2003), Dahl (2006) and Hjort, Dahl and Steinbakk (2006). There is a vast literature on other methods of evaluating *p*-values. Further discussion and references can be found in Bayarri and Berger (2000).

Here, we focus on studying the degree to which the various *p*-values deviate from uniformity in finite sample scenarios. For this purpose, we simulate the null sampling distribution of $p_{prior}^{EB}(\mathbf{X})$, $p_{post}(\mathbf{X})$ and $p_{ppp}(\mathbf{X})$, when $\mathbf{X}$ comes from a hierarchical normal-normal model as defined in (3.1). [We do not show the behavior of $p_{post}^{EB}(\mathbf{X})$ because it is basically identical to that of $p_{post}(\mathbf{X})$.]

In particular, we have simulated 1000 samples from the following model:

$$X_{ij} \sim N(\theta_i, 4), \quad i = 1, \dots, I, j = 1, \dots, 8,$$
$$\theta_i \sim N(0, 1), \quad i = 1, \dots, I.$$

We have considered three different "group sizes": $I = 5$, 15 and 25. (Since here we are checking the distribution of the means, the adequate "asymptotics" is in the number of groups.)

We compute the different *p*-values for 1000 simulated samples. Figure 1 shows the resulting histograms. As we can see, $p_{ppp}(\mathbf{X})$ has already a (nearly) uniform distribution even for $I$ (number of groups) as small as 5. On the other hand, the distributions of both $p_{prior}^{EB}(\mathbf{X})$ and $p_{post}(\mathbf{X})$ are quite far from uniformity, the distribution of $p_{post}(\mathbf{X})$ being the farthest. Moreover, the deviation from the $U(0, 1)$ is in the direction of more conservatism (given little probability to small *p*-values, and concentrating around 0.5), as it is the case in simpler models. Notice that conservatism usually results in lack of power (and thus in not being able to detect data coming from wrong models). Particularly worrisome is the behavior of $p_{post}(\mathbf{X})$ for small number of groups. We have also performed similar simulations for larger $I$'s (number of groups) to investigate whether the distribution of these $p$-values approaches uniformity as $I$ grows. In Figure 2 we show the histograms for $I = 100$ and $I = 200$ of $p$-values $p_{post}(\mathbf{X})$ and $p_{prior}^{EB}(\mathbf{X})$ [we do not show $p_{ppp}(\mathbf{X})$ as it is virtually uniform]. The distributions of these $p$-values do not seem to change much as $I$ is doubled from $I = 100$ to $I = 200$, and they are still quite far from uniformity, still showing a tendency to concentrate around middle values for $p$. We do not know whether these $p$-values are asymptotically $U(0,1)$.

### 3.6 Distribution of $p$-Values Under Some Alternatives

In this section we study the behavior of $p_{prior}^{EB}(\mathbf{X})$, $p_{post}(\mathbf{X})$ and $p_{ppp}(\mathbf{X})$, when the "null" normal-normal model is wrong. In particular, we focus on violations of normality at the second level.

Specifically, we simulate data sets from three different models. In all the three, we take the distribution at the first level to be the same and in agreement with the first level in the null model (3.1):

$$X_{ij} \sim N(\theta_i, \sigma^2 = 4), \quad i = 1, \ldots, I, j = 1, \ldots, 8.$$

We use three different distributions for the group means (remember, under the null model, the $\theta_i$'s were normal):

1. Exponential distribution: $\theta_i \sim \text{Exp}(1), i = 1, \ldots, I$.
2. Gumbel distribution: $\theta_i \sim \text{Gumbel}(0, 2), i = 1, \ldots, I$, where the Gumbel$(\alpha, \beta)$ density is

$$f(x \mid \alpha, \beta) = \frac{1}{\beta} \exp\left(-\frac{x - \alpha}{\beta}\right)$$
$$\cdot \exp\left(-\exp\left(-\frac{x - \alpha}{\beta}\right)\right),$$

   for $-\infty < x < \infty$. Gumbel distribution is also known as *Extreme Value Type I distribution*. It is skew, with a long tail to the right (left) when derived as the limiting distribution of a maximum (minimum).
3. Log-Normal distribution: $\theta_i \sim \text{LogNormal}(0, 1)$, $i = 1, \ldots, I$.

We have considered $I = 5$ and $I = 10$, simulated 1000 samples from each model and computed the different $p$-values for each sample. In Table 2 we show $Pr(p \leq \alpha)$ for the three $p$-values and some values of $\alpha$. $p_{ppp}$ seems to show decent power given the small sample sizes and number of groups (power is lower for the exponential alternative, and largest for the log-normal); both $p_{prior}^{EB}$ and $p_{post}$ show considerable lack of power in comparison. In particular, notice the extreme low power of $p_{post}$ in all instances, producing basically no $p$-values smaller than 0.2.

Table 2
$Pr(p \leq \alpha)$ for $p_{ppp}$, $p_{post}$ and $p_{prior}^{EB}$, for different values of $I$ and the three alternative models

| $\alpha$ | 0.02 | 0.05 | 0.1 | 0.2 | 0.02 | 0.05 | 0.1 | 0.2 |
|---|---|---|---|---|---|---|---|---|
| | Normal-Exponential | | | | | | | |
| | $I = 5$ | | | | $I = 10$ | | | |
| $p_{ppp}$ | 0.04 | 0.08 | 0.15 | 0.24 | 0.12 | 0.20 | 0.29 | 0.42 |
| $p_{post}$ | 0.00 | 0.00 | 0.00 | 0.00 | 0.00 | 0.00 | 0.01 | 0.05 |
| $p_{prior}^{EB}$ | 0.00 | 0.00 | 0.00 | 0.23 | 0.00 | 0.06 | 0.18 | 0.37 |
| | Normal-Gumbel | | | | | | | |
| | $I = 5$ | | | | $I = 10$ | | | |
| $p_{ppp}$ | 0.12 | 0.22 | 0.32 | 0.46 | 0.21 | 0.31 | 0.42 | 0.55 |
| $p_{post}$ | 0.00 | 0.00 | 0.00 | 0.00 | 0.00 | 0.00 | 0.00 | 0.00 |
| $p_{prior}^{EB}$ | 0.00 | 0.00 | 0.00 | 0.23 | 0.00 | 0.07 | 0.19 | 0.38 |
| | Normal-Lognormal | | | | | | | |
| | $I = 5$ | | | | $I = 10$ | | | |
| $p_{ppp}$ | 0.16 | 0.22 | 0.31 | 0.41 | 0.32 | 0.42 | 0.50 | 0.61 |
| $p_{post}$ | 0.00 | 0.00 | 0.00 | 0.00 | 0.00 | 0.00 | 0.00 | 0.02 |
| $p_{prior}^{EB}$ | 0.00 | 0.00 | 0.00 | 0.23 | 0.01 | 0.06 | 0.13 | 0.23 |

## 4. TESTING $\mu = \mu_0$

As we have seen in Section 3, the specified predictive distributions for $T$ (empirical Bayes, posterior and partial posterior) used to locate the observed $t_{obs}$ had to be dealt with by MC and MCMC methods. To gain understanding in the behavior of the different proposals to "get rid" of the unknown parameters, we address here a simpler "null model" which results in simpler expressions and allows for easier comparisons.

Suppose that we have the normal-normal hierarchical model (3.1) (with $\sigma_i^2$ known) but that we are interested in testing

$$H_0 : \mu = \mu_0.$$

A natural $T$ to consider to investigate this $H_0$ is the grand mean:

$$T = \frac{\sum_{i=1}^{I} n_i \overline{X}_{i\cdot}}{\sum_{i=1}^{I} n_i},$$

where $\overline{X}_{i\cdot}$, $i = 1, \ldots, I$, are the sample means for the $I$ groups. The (null) sampling distribution of $T$ is:

$$T \mid \boldsymbol{\theta} \sim N(\mu_T, V_T)$$

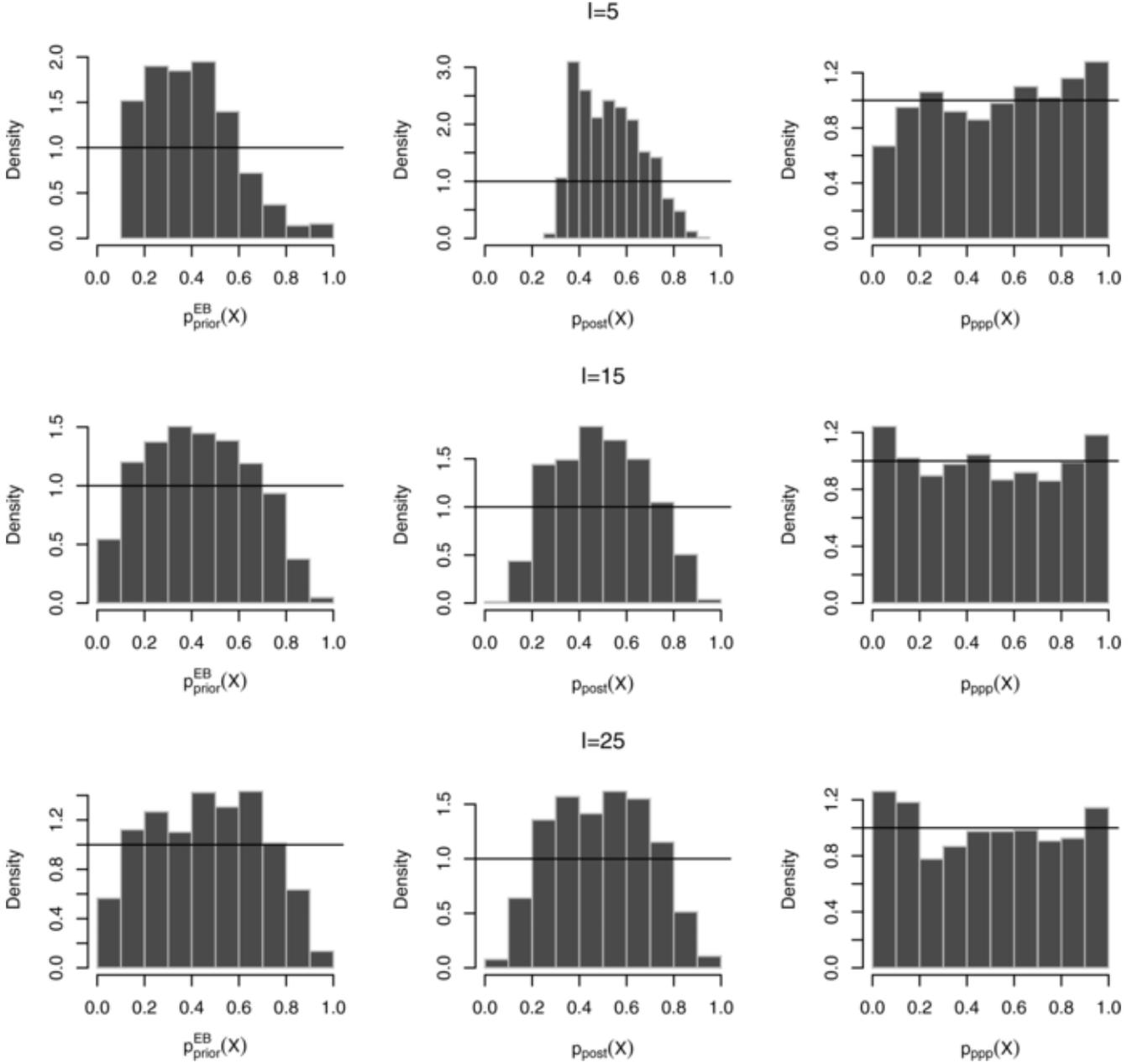

FIG. 1. *Null distribution of $p^{EB}_{prior}(\mathbf{X})$ (first column), $p_{post}(\mathbf{X})$ (second column) and $p_{ppp}(\mathbf{X})$ (third column) for $I = 5$ (first row), 15 (second row) and 25 (third row).*

$$ (4.1) \qquad \text{with } \mu_T = \frac{\sum_{i=1}^{I} n_i \theta_i}{\sum_{i=1}^{I} n_i}, V_T = \frac{\sum_{i=1}^{I} n_i \sigma_i^2}{(\sum_{i=1}^{I} n_i)^2}. $$

Again we will integrate $\boldsymbol{\theta}$ out from (4.1) with the previous proposals and compare the resulting predictive distributions for $T$, $h(t)$, and the corresponding MS (which we take relative to $\mu_0$):

$$ (4.2) \qquad p = Pr^{h(\cdot)}(|t(\mathbf{X}) - \mu_0| \geq |t(\mathbf{x}_{obs}) - \mu_0|), $$

$$ (4.3) \qquad RPS = \frac{h(t(\mathbf{x}_{obs}))/h(\mu_0)}{\sup_t h(t)/h(\mu_0)}. $$

### 4.1 Empirical Bayes Distributions

In this case the integrated likelihood for $\tau^2$ is simply given by (3.4) with $\mu$ replaced by $\mu_0$, from which $\hat{\tau}^2$ the MLE of $\tau^2$ can be computed. For this null model, it is possible to derive closed form expressions for the prior and posterior empirical Bayes distributions given in (2.2) and (2.4), respectively.

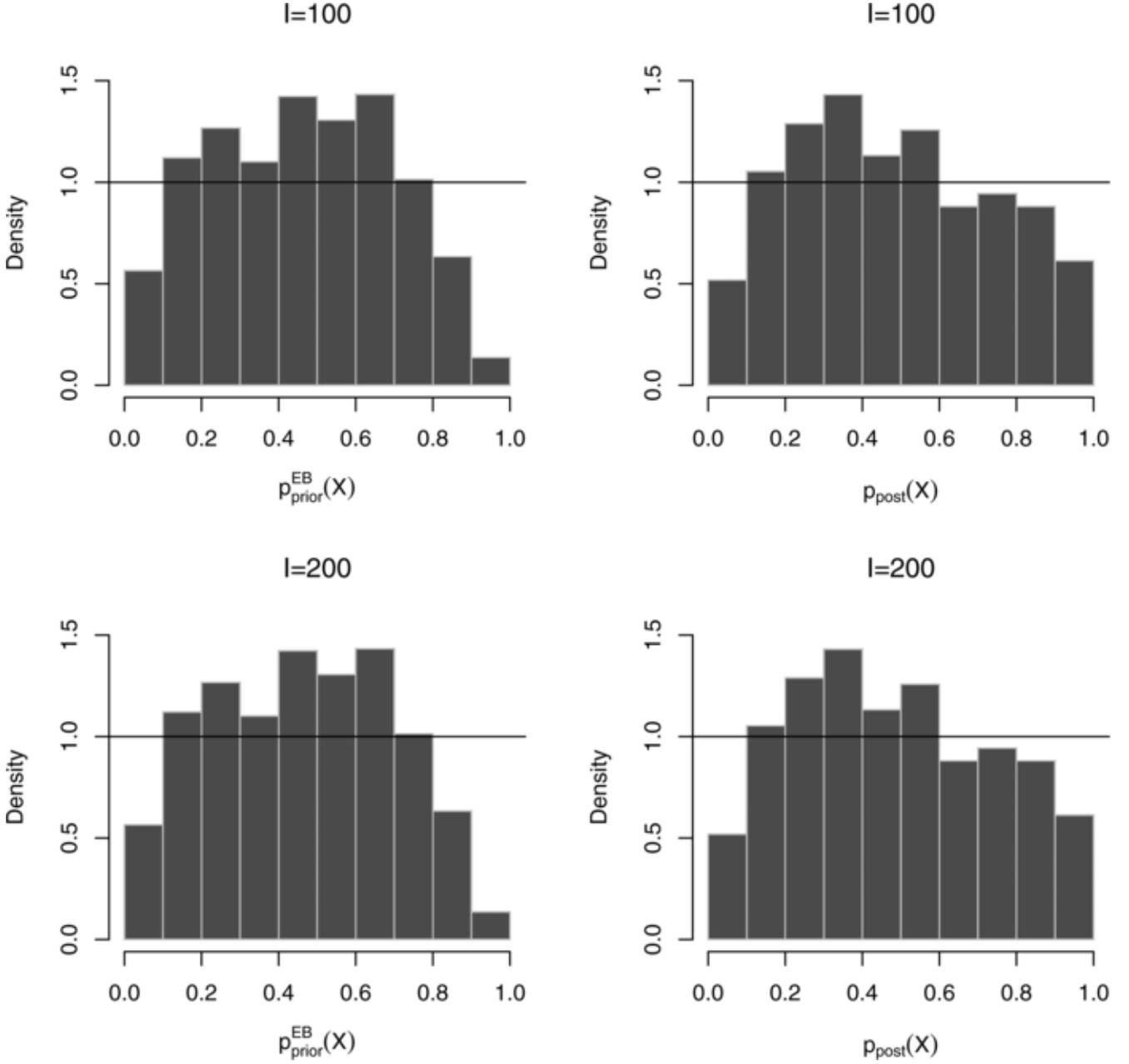

FIG. 2. *Null distribution of $p_{prior}^{EB}(\mathbf{X})$ and $p_{post}(\mathbf{X})$ when $I = 100$ (first row) and $I = 200$ (second row).*

Indeed, the joint empirical Bayes prior predictive for $\overline{\mathbf{X}} = (\overline{X}_1, \ldots, \overline{X}_I.)$ is

$$m_{prior}^{EB}(\bar{\mathbf{x}}) = \prod_{i=1}^{I} N\left(\bar{x}_i. \,\Big|\, \mu_0, \frac{\sigma_i^2}{n_i} + \hat{\tau}^2\right),$$

so that the corresponding distribution for $T$, $m_{prior}^{EB}(t)$, is normal with mean and variance given by

$$E_{prior}^{EB} = \mu_0,$$

(4.4)

$$V_{prior}^{EB} = \frac{1}{(\sum_{i=1}^{I} n_i)^2} \sum_{i=1}^{I} n_i^2 \left(\frac{\sigma_i^2}{n_i} + \hat{\tau}^2\right).$$

The empirical Bayes *posterior* predictive distribution $m_{post}^{EB}(\bar{\mathbf{x}})$ can be derived in a similar manner resulting also in a normal $m_{post}^{EB}(t)$ with mean and variance

$$E_{post}^{EB} = \frac{\sum_{i=1}^{I} n_i \widetilde{E}_i}{\sum_{i=1}^{I} n_i},$$

(4.5)
$$V_{post}^{EB} = \frac{1}{(\sum_{i=1}^{I} n_i)^2} \sum_{i=1}^{I} n_i^2 \left( \frac{\sigma_i^2}{n_i} + \widetilde{V}_i \right),$$

where
$$\widetilde{E}_i = \frac{n_i \overline{x}_{i\cdot}/\sigma_i^2 + \mu_0/\hat{\tau}^2}{n_i/\sigma_i^2 + 1/\hat{\tau}^2}$$

and
$$\widetilde{V}_i = \frac{1}{n_i/\sigma_i^2 + 1/\hat{\tau}^2}.$$

The MS (4.2) and (4.3) can also be computed in closed form. The (prior) empirical Bayes measures are

$$p_{prior}^{EB} = 2 \cdot \left( 1 - \Phi\left( \frac{|t_{obs} - \mu_0|}{\sqrt{V_{prior}^{EB}}} \right) \right),$$

$$RPS_{prior}^{EB} = \exp\left\{ -\frac{(t_{obs} - \mu_0)^2}{2V_{prior}^{EB}} \right\},$$

where $\Phi$ denotes the standard normal distribution function. The *posterior* empirical Bayes measures can similarly be derived in closed form, but they are of much less interest and we do not produce them here (see Castellanos (2002)).

The inadequacies of $m_{post}^{EB}$ for testing the null model can already be seen in the above formulas, but they are more evident in the particular homoscedastic, balanced case: $\sigma_i^2 = \sigma^2$ and $n_i = n \ \forall i, i = 1, \ldots, I$. In this case the distribution of $T$ simplifies to

$$T \sim N\left( \frac{\sum_{i=1}^{I} \theta_i}{I}, \frac{\sigma^2}{In} \right).$$

Also, there is a closed form expression for the MLE of $\tau^2$:

$$\hat{\tau}^2 = \max\left\{ 0, \frac{\sum_{i=1}^{I}(\overline{x}_{i\cdot} - \mu_0)^2}{I} - \frac{\sigma^2}{n} \right\}.$$

Then, the mean and variance of $m_{prior}^{EB}$, as given in (4.4), are

$$E_{prior}^{EB} = \mu_0, \quad V_{prior}^{EB} = \frac{\frac{\sigma^2}{n} + \hat{\tau}^2}{I}.$$

Similarly, the mean and variance of $m_{post}^{EB}$, given in (4.5), reduce to

$$E_{post}^{EB} = \frac{nt_{obs}/\sigma^2 + \mu_0/\hat{\tau}^2}{n/\sigma^2 + 1/\hat{\tau}^2},$$

$$V_{post}^{EB} = \frac{2n\sigma^2\hat{\tau}^2 + \sigma^4}{nI(n\hat{\tau}^2 + \sigma^2)}.$$

For a given $\mu_0$ (and fixed $\tau$), it is now easy to investigate the behavior of $m_{prior}^{EB}$ and $m_{post}^{EB}$ as $t_{obs} \to \infty$, indicating flagrant incompatibility between the data and $H_0$. The comparison in this simple case is enlightening. First, note that $m_{prior}^{EB}$ centers at $\mu_0$, which in principle allows for declaring incompatible a very large value $t_{obs}$; however, the variance also grows to $\infty$ as $t_{obs}$ grows, thus alleviating the incompatibility, and maybe "missing" some surprisingly large $t_{obs}$. Thus, the behavior of $m_{prior}^{EB}$ is reasonable, but might be conservative. On the other hand, the behavior of $m_{post}^{EB}$ is completely inadequate. Indeed, for very large $t_{obs}$, it centers precisely at $t_{obs}$, thus precluding detecting as unusual *any* value $t_{obs}$, no matter how large! Moreover, the variance goes to $(2\sigma^2)/(nI)$, a finite constant. It is immediate to see that $m_{post}^{EB}$ should not be used to check this particular (and admittedly simple) model; as a matter of fact, for $t_{obs} \to \infty$ (extremely inadequate models) we expect $p$-values of around 0.5. We remark that the previous argument does not belong to any particular MS; rather it reflects the inadequacy of $m_{post}^{EB}$ for model checking, whatever MS we use. Note that we expect similar inadequacies to occur with the posterior predictive distribution, which is rather often used in objective Bayes model checking.

### 4.2 Posterior Distribution

No major simplifications occur for this specific $H_0$. The posterior distribution is not of closed form (not even for the homoscedastic, balanced case), and hence neither is the posterior predictive distribution. We can, however, easily generate from it with virtually the same Gibbs sampler used in Section 3.2: it suffices to (obviously) ignore the full conditional for $\mu$ and replace $\mu$ with the value $\mu_0$ in the other two full conditionals (B.2) and (B.3), which were standard distributions.

### 4.3 Partial Posterior Distribution

There is no closed form expression for the partial posterior distribution either, but considerable simplification occurs since the Metropolis-within-Gibbs step is no longer needed and a straight Gibbs sampler suffices. The full conditional for $\tau^2$ is as given in (C.2) with $\mu$ replaced by $\mu_0$; the full conditional of each $\theta_i$ is here also normal:

$$\pi(\theta_i \mid \tau^2, \boldsymbol{\theta}_{-i}, \mathbf{x}_{obs} \setminus t_{obs}) = N(\theta_i \mid E_i^0, V_i^0),$$

where

$$E_i^0 = \frac{1}{V_i^0}\left[\frac{n_i}{\sigma_i^2}\left(\overline{x}_{i\cdot} - \frac{\sigma_i^2}{\sum_j n_j \sigma_j^2}\right.\right.$$

$$(4.6) \qquad \left.\cdot\left(\sum_j n_j t_{obs} - \sum_{l \neq i} n_l \theta_l\right)\right)$$

$$\left.+ \frac{1}{\tau^2}\mu_0\right],$$

$$(4.7) \qquad \frac{1}{V_i^0} = \frac{n_i}{\sigma_i^2}\left(1 - \frac{n_i\sigma_i^2}{\sum_{j=1}^{I} n_j\sigma_j^2}\right) + \frac{1}{\tau^2}.$$

Details of the derivations appear in Appendix D.

### 4.4 Some Examples

We next consider four examples in which we carry out the testing $H_0 : \mu = 0$. We consider $I = 8$ groups, with $n = 12$ observations per group, and $\sigma^2 = 4$. In one of the examples (Example 1) $H_0$ is true and the means $\theta_i$ are generated from a $N(0,1)$. In the remaining three examples the null $H_0$ is wrong, with $\theta_i \sim N(1.5,1)$ in Example 2, $\theta_i \sim N(2.5,1)$ in Example 3, and $\theta_i \sim N(2.5,3)$ in Example 4. The simulated sample means are:

EXAMPLE 3.

$$\overline{\mathbf{x}} = (-2.18, -1.47, -0.87, -0.38,$$
$$0.05, 0.29, 0.96, 2.74).$$

EXAMPLE 4.

$$\overline{\mathbf{x}} = (-0.05, 0.66, 1.37, 1.70, 1.72, 2.14, 2.73, 3.68).$$

EXAMPLE 5.

$$\overline{\mathbf{x}} = (1.53, 1.65, 1.71, 1.75, 1.87, 2.16, 2.47, 3.68).$$

EXAMPLE 6.

$$\overline{\mathbf{x}} = (0.50, 1.52, 1.59, 2.73, 2.88, 3.54, 4.21, 5.86).$$

In Figure 3 we show the predictive distributions for all proposals in the four examples. A quite remarkable feature is that in every occasion, $m_{post}^{EB}$ basically coincides with $m_{post}$, so much that they can hardly be told apart. We were expecting them to be close, but not so close. Also, when the null is true (Example 1), all distributions rightly concentrate around the null and, as expected, the most concentrated is $m_{post}^{EB}$ (and $m_{post}$), and the least is $m_{ppp}$ ($m_{prior}^{EB}$ ignores the uncertainty in the estimation of $\tau^2$). When the null model is wrong, however, even though both $m_{ppp}$ and $m_{prior}^{EB}$ have the right lo-cation, $m_{ppp}$ is more concentrated than $m_{prior}^{EB}$, thus indicating more promise in detecting extreme $t_{obs}$. Notice the hopeless (and identical) behavior of $m_{post}^{EB}$ and $m_{post}$: both concentrate around $t_{obs}$, no matter how extreme; that is, there is no hope that it can detect incompatibility of a very large $t_{obs}$ with the hypothetical value of 0.

In Table 3 we show the different MS for the four examples. All behave well when the null is true, but only the ppp and the prior empirical Bayes measures detect the wrong models (ppp more clearly). On the other hand, $m_{post}^{EB}$ and $m_{post}$ produce very similar measures and both are incapable of detecting clearly inappropriate null models. Notice that the conservatism of the posterior predictive measures (and the posterior empirical Bayes ones) is extreme.

## 5. A COMPARISON WITH OTHER BAYESIAN METHODS

In this section we retake the main goal of checking the adequacy of the second level in the hierarchical model:

$$X_{ij} \mid \theta_i \overset{i}{\sim} N(\theta_i, \sigma^2),$$
$$i = 1, \ldots, I, j = 1, \ldots, n_i,$$
$$\pi(\boldsymbol{\theta} \mid \mu, \tau) = \prod_{i=1}^{I} N(\theta_i \mid \mu, \tau^2),$$

with $\sigma^2$ unknown, as well as $\mu, \tau^2$. We first provide some details needed to derive the MS used so far when $\sigma^2$ is unknown; we then briefly review three recent methods for Bayesian checking of hierarchical models, proposed in Dey, Gelfand, Swartz and Vlachos (1998), O'Hagan (2003) and Marshall and Spiegelhalter (2003). We do not specifically address here (because the philosophy is somewhat different) the much earlier, likelihood/empirical Bayes proposal of Lange and Ryan (1989), which basically consists in checking the normality of some standardized version of estimated residuals. We apply the four methods considered so far and the three new methods to a data set proposed in O'Hagan (2003).

TABLE 3
*p-values and RPS for testing $\mu = 0$ in the four examples*

|  | Example 1 | | Example 2 | | Example 3 | | Example 4 | |
|---|---|---|---|---|---|---|---|---|
|  | p | RPS | p | RPS | p | RPS | p | RPS |
| ppp | 0.86 | 0.98 | 0.01 | 0.01 | 0.00 | 0.00 | 0.00 | 0.01 |
| EB prior | 0.83 | 0.98 | 0.02 | 0.06 | 0.01 | 0.03 | 0.01 | 0.05 |
| EB post | 0.71 | 1.00 | 0.31 | 0.89 | 0.30 | 0.88 | 0.38 | 1.00 |
| post | 0.71 | 1.00 | 0.33 | 0.92 | 0.32 | 0.95 | 0.39 | 1.00 |

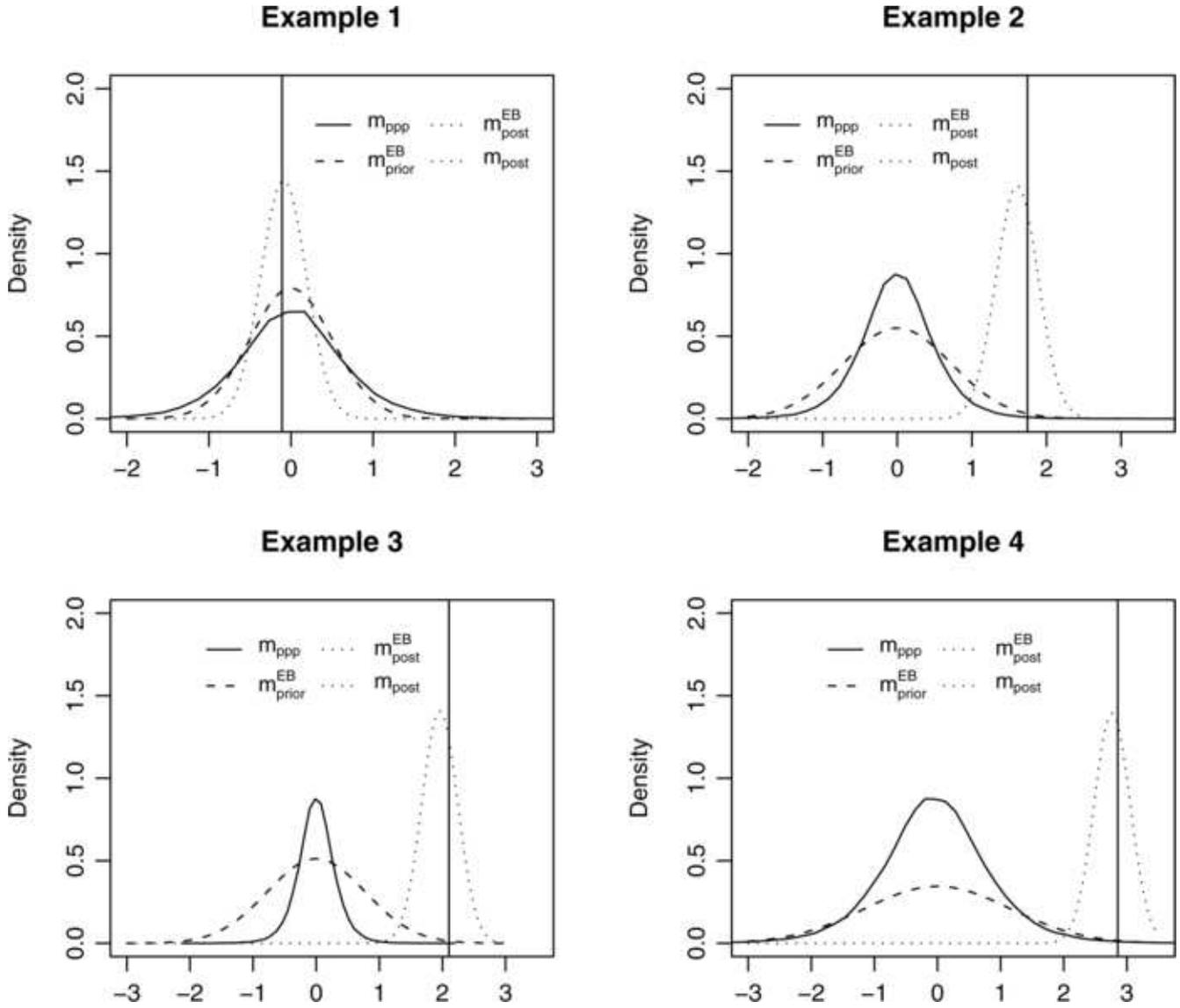

FIG. 3. *Different predictive distribution for T in each example. The vertical solid line locates $t_{obs}$. The curves corresponding to $m_{post}$ and $m_{post}^{EB}$ were almost indistinguishable and for clarity are represented as identical.*

O'HAGAN (2003) EXAMPLE.   In the general scenario of checking the normal-normal hierarchical model, O'Hagan (2003) uses the following data set:

| Group 1 | 2.73, | 0.56, | 0.87, | 0.90, | 2.27, | 0.82. | $\overline{x}_{1\cdot} = 1.36.$ |
| Group 2 | 1.60, | 2.17, | 1.78, | 1.84, | 1.83, | 0.80. | $\overline{x}_{2\cdot} = 1.67.$ |
| Group 3 | 1.62, | 0.19, | 4.10, | 0.65, | 1.98, | 0.86. | $\overline{x}_{3\cdot} = 1.57.$ |
| Group 4 | 0.96, | 1.92, | 0.96, | 1.83, | 0.94, | 1.42. | $\overline{x}_{4\cdot} = 1.34.$ |
| Group 5 | 6.32, | 3.66, | 4.51, | 3.29, | 5.61, | 3.27. | $\overline{x}_{5\cdot} = 4.44.$ |

Note that $\overline{x}_{5\cdot}$ is considerably far from the other four sample means.

## 5.1 Methods Used So Far

The empirical Bayes methods (both the prior and the posterior) have an easy generalization to the unknown $\sigma^2$ case. It suffices to substitute $\sigma^2$ by its usual MLE estimate and apply the procedures in Section 3 for $\sigma^2$ known.

For both the posterior predictive and the partial posterior predictive measures, we need to specify a new (noninformative) joint prior. Since we can use the standard noninformative prior for $\sigma^2$, we take

$$(5.1) \qquad \pi(\mu, \sigma^2, \tau^2) \propto \frac{1}{\sigma^2} \frac{1}{\tau}.$$

To simulate from the posterior distribution, we again use Gibbs sampling. The full conditionals for $\boldsymbol{\theta}$, $\mu$ and $\tau^2$ are the same as for the known $\sigma^2$ and they are given in (B.3), (B.1) and (B.2), respectively. The full conditional for the new parameter, $\sigma^2$, is

$$\sigma^2 \mid \boldsymbol{\theta}, \mu, \tau^2, \mathbf{x}_{obs} \sim \chi^{-2}(m, \tilde{\sigma}^2),$$

where $m = \sum_{i=1}^{I} n_i$ and $\tilde{\sigma}^2 = \sum_{i=1}^{I} \sum_{j=1}^{n_i} (x_{ij} - \theta_i)^2 / n$.

The (joint) partial posterior distribution is

$$\pi_{ppp}(\boldsymbol{\theta}, \sigma^2, \mu, \tau^2 \mid \mathbf{x}_{obs} \setminus t_{obs}) \propto \frac{\pi(\boldsymbol{\theta}, \sigma^2, \mu, \tau^2 \mid \mathbf{x}_{obs})}{f(t_{obs} \mid \boldsymbol{\theta}, \sigma^2)},$$

and again we use the same general procedure as for the $\sigma^2$ known scenario (see Section 3). We only need to specify how to simulate from the full conditional of $\sigma^2$:

$$\pi_{ppp}(\sigma^2 \mid \boldsymbol{\theta}, \mu, \tau^2, \mathbf{x}_{obs} \setminus t_{obs}) \propto \frac{\chi^{-2}(m, \tilde{\sigma}^2)}{f(t_{obs} \mid \boldsymbol{\theta}, \sigma^2)}.$$

We use Metropolis–Hastings with $\chi^{-2}(m, \tilde{\sigma}^2)$ as proposal distribution. The acceptance probability (at stage $k$) of candidate $\sigma^{2*}$, given the simulated values $(\boldsymbol{\theta}^{(k)}, \sigma^{2(k)}, \mu^{(k)}, \tau^{2(k)})$, is

$$\alpha = \min\left\{ 1, \frac{f(t_{obs} \mid \boldsymbol{\theta}^{(k)}, \sigma^{2(k)})}{f(t_{obs} \mid \boldsymbol{\theta}^{(k)}, \sigma^{2*})} \right\}.$$

We next derive the different MS for O'Hagan data.

O'HAGAN (2003) EXAMPLE (CONTINUED). The empirical Bayes, posterior predictive and partial posterior predictive MS applied to this data set, using $T = \max_i\{\overline{X}_i\}$, are shown in Table 4.

We again observe the same behavior as the one repeatedly observed in previous examples: in spite of such an "obvious" data set, only the partial posterior measures detect the incompatibility between data and model. The empirical Bayes prior measures are too conservative, and the posterior predictive measures (and their very much alike empirical Bayes posterior ones) are completely hopeless.

## 5.2 Simulation-Based Model Checking

This method is proposed in Dey, Gelfand, Swartz and Vlachos (1998), as a computationally intense method for model checking.

This method works not only with checking statistics $T$, but more generally, with discrepancy measures $d$, that is, with functions of the parameters and the data. This feature also applies to the posterior predictive checks that we have been considering all along. In essence, the method consists in comparing the posterior distribution $d \mid \mathbf{x}_{obs}$ with $R$ posterior distributions of $d$ given $R$ data sets $\mathbf{x}^r$, for $r = 1, \ldots, R$, generated from the (null) *prior predictive* model. Note that this method *requires* proper priors. Comparison is carried out via *Monte Carlo Tests* (Besag and Clifford (1989)).

Letting $\mathbf{x}^r$, for $r = 0$, denote the observed data $\mathbf{x}_{obs}$, their algorithm is as follows:

- For each posterior distribution of $d$ given $\mathbf{x}^r, r = 0, \ldots, R$, compute the vector of quantiles $\mathbf{q}^{(r)} = (q_{0.05}^{(r)}, q_{0.25}^{(r)}, q_{0.5}^{(r)}, q_{0.75}^{(r)}, q_{0.95}^{(r)})$, where $q_\alpha^{(r)}$ is the $\alpha$-quantile of the posterior distribution given data $\mathbf{x}^r, r = 0, \ldots, R$.
- Compute the vector $\overline{\mathbf{q}}$ of averages, over $r$, of these quantiles: $\overline{\mathbf{q}} = (\overline{q}_{0.05}, \overline{q}_{0.25}, \overline{q}_{0.5}, \overline{q}_{0.75}, \overline{q}_{0.95})$.
- Compute the $r + 1$ Euclidean distances between $\mathbf{q}^{(r)}, r = 0, 1, \ldots, R$ and $\overline{\mathbf{q}}$.
- Perform a 0.05 one-sided, upper tail Monte Carlo test, that is, check whether or not the distance corresponding to the original data is smaller than the 95th percentile of the $r + 1$ distances.

In reality, this method is not a competitor of the ones we have been considering previously, since it *requires* proper priors, and hence is not available for objective model checking. We, however, apply it also to O'Hagan data.

O'HAGAN (2003) EXAMPLE (CONTINUED). In order to perform the *simulation-based model checking*, we need proper priors. We use the ones proposed in O'Hagan (2003):

$$(5.2) \qquad \mu \sim N(2, 10), \quad \sigma^2 \sim 22W, \quad \tau^2 \sim 6W$$

$$\text{where } W \sim \chi_{20}^{-2}.$$

Along the statistic used so far, we have also considered a measure of discrepancy which in this case is just a function of the parameters:

$$T_1 = \max \overline{X}_i., \quad T_2 = \max |\theta_i - \mu|.$$

With 1000 simulated data sets from the null, the results are shown in Table 5. It can be seen that, with the given prior, incompatibility is detected with $T_2$, but not with $T_1$. We do not know whether $T_2$ would detect incompatibility with other priors (see related results in Section 5.3).

TABLE 4
*MS ($\sigma^2$ unknown) for O'Hagan data set*

| $p_{prior}^{EB}$ | $RPS_{prior}^{EB}$ | $p_{post}^{EB}$ | $RPS_{post}^{EB}$ | $p_{post}$ | $RPS_{post}$ | $p_{ppp}$ | $RPS_{ppp}$ |
|---|---|---|---|---|---|---|---|
| 0.19 | 0.4 | 0.37 | 0.95 | 0.40 | 0.99 | 0.01 | 0.02 |

### 5.3 O'Hagan Method

O'Hagan (2003) proposes a general method to investigate adequacy of graphical models at each node. We will not describe his method in full generality, but only how it applies to checking the second level of our normal-normal hierarchical model.

To investigate conflict between the data and the normal assumption for each of the group means, this proposal investigates conflict between the likelihood for $\theta_i$, $\prod_{j=1}^{n_i} f(x_{ij} \mid \theta_i, \sigma^2)$, and the (null) density for $\theta_i$, $\pi(\theta_i \mid \mu, \tau^2)$.

To check conflict between two known univariate densities/likelihoods, O'Hagan proposes a "measure of conflict" based on their relative heights at an "intermediate" value. Specifically, the likelihoods/densities are first normalized so that their maximum height is 1 (notice that this is equivalent to dividing by their respective maximum, as in $RPS$ before). Then the (common) density height, $z$, at the value of $\theta_i$ between the two modes where the two densities are equal, is computed. The proposed measure of conflict is $c = -2 \ln z$. For the particular case of comparing two normal distributions, $N(\omega_i, \gamma_i^2)$, for $i = 1, 2$, this measure is

$$(5.3) \qquad c = \left( \frac{\omega_1 - \omega_2}{\sqrt{\gamma_1} + \sqrt{\gamma_2}} \right)^2.$$

O'Hagan indicates that a conflict measure smaller than 1 should be taken as indicative of no conflict, whereas values of 4 or larger would indicate clear conflict. No indication is given for values lying between 1 and 4.

When, as usual, the distributions involved depend on unknown parameters, the measures of conflict [in particular (5.3)], cannot be computed. O'Hagan's proposal is then to use the median of their *posterior* distribution. Notice that this is closely related to computing a relative height on the posterior predictive distribution and, hence, the concern exists that it can be too conservative for useful model checking. In fact this conservatism was highlighted in the discussions by Bayarri (2003) and Gelfand (2003).

TABLE 5
*Euclidean distance between $\mathbf{q}^{(0)}$ and $\overline{\mathbf{q}}$ and the 0.95 quantile of all distances*

| | $\|\mathbf{q}^{(0)} - \overline{\mathbf{q}}\|$ | 0.95 quantile |
|---|---|---|
| $T_1$ | 2.31 | 13.46 |
| $T_2$ | 1.82 | 0.81 |

TABLE 6
*Posterior medians of $c_i$, $i = 1, \ldots, 5$, for O'Hagan data set*

| | $\theta_1$ | $\theta_2$ | $\theta_3$ | $\theta_4$ | $\theta_5$ |
|---|---|---|---|---|---|
| O'Hagan priors | 0.43 | 0.14 | 0.22 | 0.46 | 4.81 |
| Noninformative priors | 0.16 | 0.09 | 0.11 | 0.16 | 1.36 |

Interestingly enough, O'Hagan defends use of *proper* priors for the unknown parameters, so neither posterior predictive nor posterior distributions are needed for implementation of his proposal (since the prior predictives and priors are proper). Alternatively, if one wishes to insist on using posterior distributions (instead of the, more natural, prior distributions), then proper priors are no longer needed, and the method can thus be generalized. Accordingly, we also apply his proposal with the noninformative prior (5.1).

O'HAGAN (2003) EXAMPLE (CONTINUED). We compute the measure (5.3) for the data set proposed by O'Hagan (2003). To derive the posterior distributions, we use both the proper priors proposed by O'Hagan for this example, given in (5.2), and the noninformative prior (5.1). The posterior medians for $c$ are shown in Table 6. It can be seen that the results are very dependent on the prior used: the spurious group 5 is detected with the specific proper prior used, but not with the noninformative priors (thus suffering from the expected conservatism). We recall that data were clearly indicating an anomalous group 5.

### 5.4 "Conflict" *p*-Value

Marshall and Spiegelhalter (2003) proposed this approach based on, and generalizing, cross-validation methods (see Gelfand, Dey and Chang (1992); Bernardo and Smith (1994), Chapter 6).

In cross-validation, to check adequacy of group $i$, data in group $i$, $\mathbf{X}_i$, are used to compute the "surprise" statistic (or diagnostic measure), whereas the rest of the data, $\mathbf{X}_{-i}$, are used to train the improper prior. A *mixed* $p$-value is accordingly computed as

$$(5.4) \quad p_{i,mix} = Pr^{m_{cross}(\cdot \mid \mathbf{X}_{-i})}(T_i \geq T_i^{obs}),$$

where the completely specified distribution used to compute the $i$th $p$-value is

$$m_{cross}(t_i \mid \mathbf{X}_{-i})$$
$$= \int f(t_i \mid \theta_i, \sigma^2)\pi(\theta_i \mid \mu, \tau^2)\pi(\mu, \tau^2, \sigma^2 \mid \mathbf{X}_{-i})\,d\boldsymbol{\theta},$$

and thus there is no double use of the data.

Marshall and Spiegelhalter (2003) aim to preserve the cross-validation spirit while avoiding choice of a particular statistic or discrepancy measure $T_i = T(\mathbf{X}_i)$. Specifically, they propose use of *conflict* $p$-values for each group $i$, computed as follows:

- Simulate $\theta_i^{rep}$ from the posterior $\theta_i \mid \mathbf{X}_{-i}$.
- Simulate $\theta_i^{fix}$ from the posterior $\theta_i \mid \mathbf{X}_i$.
- Compute $\theta_i^{diff} = \theta_i^{rep} - \theta_i^{fix}$.
- Compute the "conflict" $p$-value for group $i$, $i = 1, \ldots, I$, as

$$(5.5) \quad p_{i,con} = Pr(\theta_i^{diff} \leq 0 \mid \mathbf{x}).$$

Marshall and Spiegelhalter (2003) show that for location parameters $\theta_i$, the conflict $p$-value (5.5) is equal to the cross-validation $p$-value (5.4) based on statistics $\hat{\theta}_i$ with symmetric likelihoods and using uniform priors in the derivation of $\theta_i^{fix}$.

A clear disadvantage of this approach (as well as with the cross-validation mixed $p$-values) is that we have as many $p$-values as groups, and multiplicity might be an issue. (O'Hagan's measures might suffer from it too.) Since we are dealing with $p$-values, adjustment is most likely done by classical methods [controlling either the family-wise error rate, as the Bonferroni method, or the false discovery rate and related methods, as the Benjamini and Hochberg (1995) method]. None of these methods is foolproof and the danger exists that they also result in a lack of power.

O'HAGAN (2003) EXAMPLE (CONTINUED). We compute the *conflict* $p$-values for the O'Hagan data set. We again use both, O'Hagan priors and noninformative priors. The results are shown in Table 7. Taken at face value, these $p$-values behave nicely and detect the outlying group.

TABLE 7
*Conflict p-values for the O'Hagan data set using noninformative priors and O'Hagan priors*

| | Group 1 | Group 2 | Group 3 | Group 4 | Group 5 |
|---|---|---|---|---|---|
| O'Hagan priors | 0.84 | 0.74 | 0.73 | 0.88 | 0.00 |
| Noninformative | 0.66 | 0.59 | 0.61 | 0.68 | 0.00 |

## 6. A BINOMIAL-BETA EXAMPLE: BRISTOL ROYAL INFIRMARY INQUIRY DATA

We finish the paper with a real example and a different hierarchical model. Specifically, we exemplify the different checking procedures in a hierarchical Binomial-Beta model on a data set analyzed at length in Spiegelhalter et al. (2002). Data consist in the number $n_i$ of open-heart operations and the corresponding number $Y_i$ of deaths for children under one year of age carried out in 12 hospitals in England. Data are shown in Figure 4.

We consider the following model:

$$Y_i \mid \theta_i \stackrel{i}{\sim} \text{Bin}(\theta_i, n_i), \quad i = 1, \ldots, I,$$

$$\pi(\boldsymbol{\theta} \mid \alpha, \beta) = \prod_{i=1}^{I} \text{Beta}(\theta_i \mid \alpha, \beta),$$

$$(6.1) \quad \pi(\alpha, \beta) \propto [(\psi_1(\alpha) - \psi_1(\alpha + \beta))$$
$$\cdot (\psi_1(\beta) - \psi_1(\alpha + \beta))$$
$$- \psi_1(\alpha + \beta)^2]^{1/2},$$

where $\pi(\alpha, \beta)$ is the Jeffreys prior (Yang and Berger (1997)), and $\psi_1(x) = \sum_{i=1}^{\infty}(x+i)^{-2}$ denotes the trigamma function. We use both the maximum and the minimum of the frequencies of deaths, $y_i/n_i$, as checking statistics. Also, when simulating from the partial distributions we have used the normal approximation to the binomial, $y_i/n_i \approx N(\theta_i, \theta_i(1 - \theta_i)/n_i)$, so that the conditional distribution of the maximum and the minimum has an easy closed form expression.

We compute the overall partial and posterior predictive $p$-values, and also the individual (one for each hospital) O'Hagan's conflict measures and Marshall and Spiegelhalter's conflict $p$-values. All require MCMC. We use 30,000 simulations after a warm-up period of 10,000. Algorithms in R are available in http://bayes.escet.urjc.es/˜mecastellanos/FunctionsBristol.zip.

TABLE 8
*p-values for the mortality in pediatric cardiac surgery*

| | $p_{prior}^{EB}$ | $p_{post}^{EB}$ | $p_{post}$ | $p_{ppp}$ |
|---|---|---|---|---|
| Maximum | 0.03 | 0.16 | 0.23 | 0.00 |
| Minimum | 0.67 | 0.56 | 0.62 | 0.64 |

The overall $p$-values (EB prior, EB posterior, posterior and partial posterior) appear in Table 8. Also, in Figures 5 and 6 we show the corresponding predictive distributions for, respectively, the maximum and the minimum. Both the figures and the table show that the observed minimum is well supported by the assumed models with any of the $p$-values used. However, with the maximum, the EB prior and partial posterior show incompatibility (with the $ppp$ showing more incompatibility than the EB prior), while the EB posterior and posterior $p$-values fail to do so.

The multiple conflict measures are in Table 9, and the multiple conflict $p$-values in Table 10. In these tables, "1" refers to the hospital with the lowest mortality rate, and "10" to the one with the largest. According to O'Hagan's prescriptions, no hospitals show clear indication of incompatibility; all but Bristol are compatible. On the other hand, the multiple conflict $p$-values isolates Bristol as the only one incompatible. No correction for multiplicity has been used.

## 7. CONCLUSIONS

In this paper we have investigated the checking of hierarchical models from an objective Bayesian point of view (i.e., introducing only the information in the data and model). We have explored several

TABLE 9
*Posterior medians of $c_i$, $i = 1, \ldots, 12$, for Bristol data set*

| 1 | 2 | 3 | 4 | 5 | 6 | 7 | 8 | 9 | 10 | 11 | 12 |
|---|---|---|---|---|---|---|---|---|----|----|----|
| 0.51 | 0.09 | 0.07 | 0.06 | 0.06 | 0.05 | 0.05 | 0.05 | 0.10 | 0.19 | 0.64 | 3.11 |

Hospitals are ordered from lowest to largest mortality rate.

TABLE 10
*Conflict $p$-values for each hospital*

| 1 | 2 | 3 | 4 | 5 | 6 | 7 | 8 | 9 | 10 | 11 | 12 |
|---|---|---|---|---|---|---|---|---|----|----|----|
| 0.89 | 0.72 | 0.70 | 0.71 | 0.70 | 0.66 | 0.46 | 0.47 | 0.42 | 0.35 | 0.17 | 0.00 |

Hospitals are ordered from lowest to largest mortality rate.

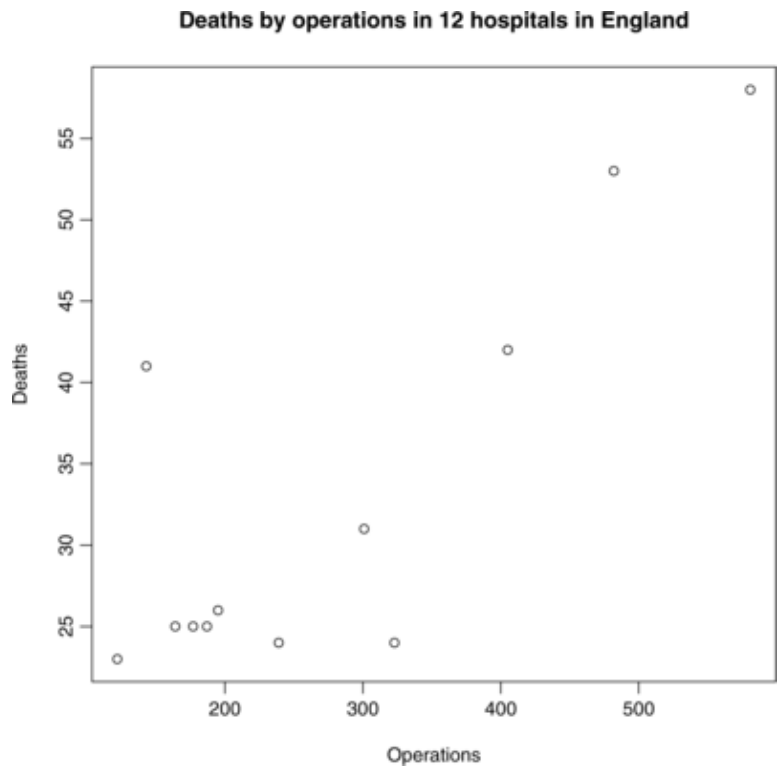

FIG. 4.   *Number of open-heart operations and deaths for children under one year of age carried out in 12 hospitals in England between 1991 and 1995.*

ways of eliminating the unknown parameters to derive "reference" distributions. We have also explored different ways of characterizing "incompatibility." We propose use of the *partial posterior predictive measures* ($MS_{ppp}$), which we compare with many other proposals. Some of our findings are:

- $MS_{ppp}$ behave considerably better than the alternative $MS_{prior}^{EB}$, $MS_{post}^{EB}$ and $MS_{post}$. The behavior of $MS_{post}$ can be particularly bad with casually chosen $T$'s, failing to reject clearly wrong models (but notice that the specific $T$ we use is the one proposed in Gelman, Carlin, Stern and Rubin (2003), Section 6.8). As a matter of fact, the measures $MS_{post}$ are very similar to the clearly inappropriate $MS_{post}^{EB}$.
- In our (limited) simulation study, the null sampling distribution of $p_{ppp}$ is found to be approximately uniform, while those of $p_{prior}^{EB}$ and $p_{post}$ are far from uniformity. Also, $p_{ppp}$ is the most powerful for the considered alternatives.
- The simulation-based model checking seems to work well in detecting the incompatibility between the model and the data, but it requires proper priors.
- The O'Hagan method is highly sensitive to the prior chosen, and in fact it seems to be conservative with noninformative priors.
- The conflict $p$-values $p_{i,con}$ seem to work well, but they produce as many $p$-values as number of groups and multiplicity might be an issue. Also, the resulting $p$-values will typically be highly dependent (any two $p$-values are based in the same data except for two observations).

Partial posterior $p$-values are not as easy to compute as posterior $p$-values, but they are still relatively easy, and indeed nothing more sophisticated than R was needed for the computations in this paper. This, along with their good properties (as demonstrated along the paper), makes them the clearly recommended procedure for objective model checking when the testing statistic $T$ is not (nearly) ancillary. But if computation is perceived as an overwhelming reason in favor of posterior $p$-values, we recommend instead use of the EB-prior $p$-values: they have better properties and are easier to compute.

## APPENDIX A: MC COMPUTATIONS FOR SECTION 3.1

To simulate from the empirical Bayes prior predictive distribution $m_{prior}^{EB}(\mathbf{x})$ simply proceed as follows: For $l = 1, \ldots, M$ simulate

$$\boldsymbol{\theta}_{(l)} = (\theta_{1(l)}, \ldots, \theta_{I(l)}) \sim \pi^{EB}(\boldsymbol{\theta}) = \prod_{i=1}^{I} \pi(\theta_i \mid \hat{\mu}, \hat{\tau}^2),$$

and for each $\boldsymbol{\theta}_{(l)}, l = 1, \ldots, M$, simulate

$$\bar{\mathbf{x}}_{(l)} = (\overline{x}_{1 \cdot (l)}, \ldots, \overline{x}_{I \cdot (l)})$$
$$\sim f(\bar{\mathbf{x}} \mid \boldsymbol{\theta}_{(l)}) = \prod_{i=1}^{I} f(\overline{x}_{i \cdot} \mid \theta_{i(l)}).$$

Simulations for the empirical Bayes posterior predictive $m_{post}^{EB}(\mathbf{x})$ proceed along the same lines except that $\theta$ is now simulated from

$$\boldsymbol{\theta}_{(l)} = (\theta_{1(l)}, \ldots, \theta_{I(l)}) \sim \pi^{EB}(\boldsymbol{\theta} \mid \mathbf{x}_{obs}) = \prod_{i=1}^{I} N(\widehat{E}_i, \widehat{V}_i),$$

where

$$\widehat{E}_i = \frac{n_i \overline{x}_{i \cdot} / \sigma_i^2 + \hat{\mu} / \hat{\tau}^2}{n_i / \sigma_i^2 + 1/\hat{\tau}^2}$$

and

$$\widehat{V}_i = \frac{1}{n_i / \sigma_i^2 + 1/\hat{\tau}^2}.$$

## APPENDIX B: FULL CONDITIONAL FOR THE GIBBS SAMPLER IN SECTION 3.2

To simulate from the joint posterior (3.5) we use an easy Gibbs sampler defined by the full conditionals

(B.1)
$$\mu \mid \boldsymbol{\theta}, \tau^2, \mathbf{x}_{obs} \sim N(E_\mu, V_\mu)$$
$$\text{with } E_\mu = \frac{\sum_{i=1}^{I} \theta_i}{I} \text{ and } V_\mu = \frac{\tau^2}{I},$$

(B.2)
$$\tau^2 \mid \boldsymbol{\theta}, \mu, \mathbf{x}_{obs} \sim \chi^{-2}(I-1, \tilde{\tau}^2)$$
$$\text{where } \tilde{\tau}^2 = \frac{\sum_{i=1}^{I} (\theta_i - \mu)^2}{I-1},$$

(B.3)
$$\theta_i \mid \mu, \tau^2, \mathbf{x}_{obs} \sim N(E_i, V_i), \text{where}$$
$$E_i = \frac{n_i \overline{x}_{i \cdot} / \sigma_i^2 + \mu / \tau^2}{n_i / \sigma_i^2 + 1/\tau^2} \text{ and } V_i = \frac{1}{n_i / \sigma_i^2 + 1/\tau^2}.$$

All the full conditionals are standard distributions, trivial to simulate from. $\chi^{-2}(\nu, a)$ refers to a *scaled* inverse chi-square distribution: it is the distribution of $(\nu a)/Y$ where $Y \sim \chi^2(\nu)$.

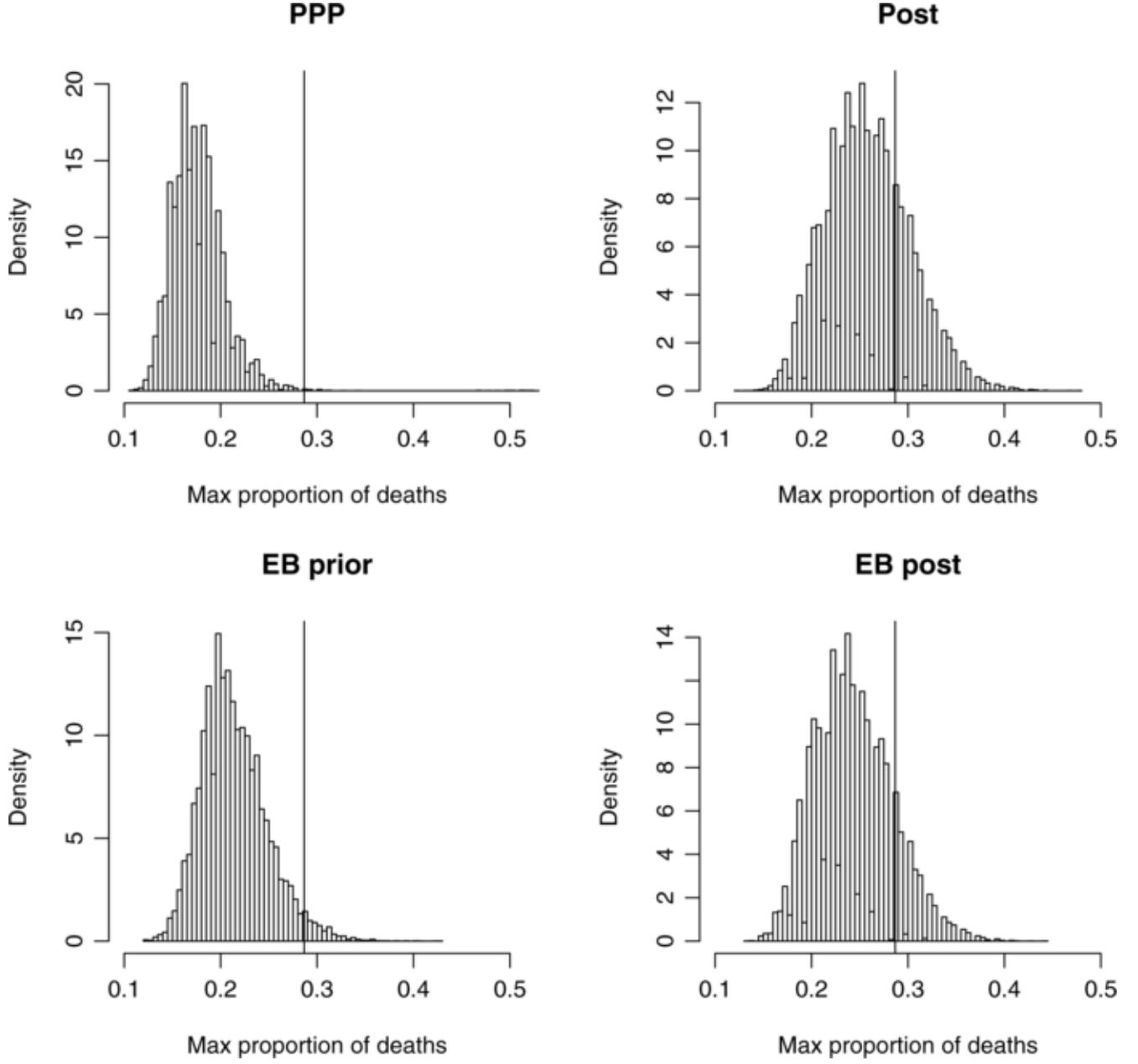

Fig. 5.   *Predictive distribution for $T = \max\{y_i/n_i\}$ in the Bristol Royal Infirmary data.*

## APPENDIX C: DETAILS FOR MCMC COMPUTATIONS IN SECTION 3.3

The full conditionals for the Gibbs sampler are

$$(C.1) \quad \mu \mid \boldsymbol{\theta}, \tau^2, \mathbf{x}_{obs} \setminus t_{obs} \propto \frac{\pi(\mu \mid \boldsymbol{\theta}, \tau^2, \mathbf{x}_{obs})}{f(t_{obs} \mid \boldsymbol{\theta})}$$
$$\propto \pi(\mu \mid \boldsymbol{\theta}, \tau^2, \mathbf{x}_{obs}),$$

$$\tau^2 \mid \boldsymbol{\theta}, \mu, \mathbf{x}_{obs} \setminus t_{obs} \propto \frac{\pi(\tau^2 \mid \boldsymbol{\theta}, \mu, \mathbf{x}_{obs})}{f(t_{obs} \mid \boldsymbol{\theta})}$$

$$(C.2) \qquad \qquad \propto \pi(\tau^2 \mid \boldsymbol{\theta}, \mu, \mathbf{x}_{obs}),$$

$$(C.3) \quad \boldsymbol{\theta} \mid \mu, \tau^2, \mathbf{x}_{obs} \setminus t_{obs} \propto \frac{\pi(\boldsymbol{\theta} \mid \mu, \tau^2, \mathbf{x}_{obs})}{f(t_{obs} \mid \boldsymbol{\theta})}.$$

The full conditionals (C.1) and (C.2) are identical to (B.1) and (B.2), respectively, and hence they are easy to simulate from. Equation (C.3) is not of closed form, and we use Metropolis–Hastings within Gibbs for the full conditional of each $\theta_i$:

$$\pi_{ppp}(\theta_i \mid \mu, \tau, \boldsymbol{\theta}_{-i}, \mathbf{x}_{obs} \setminus t_{obs})$$

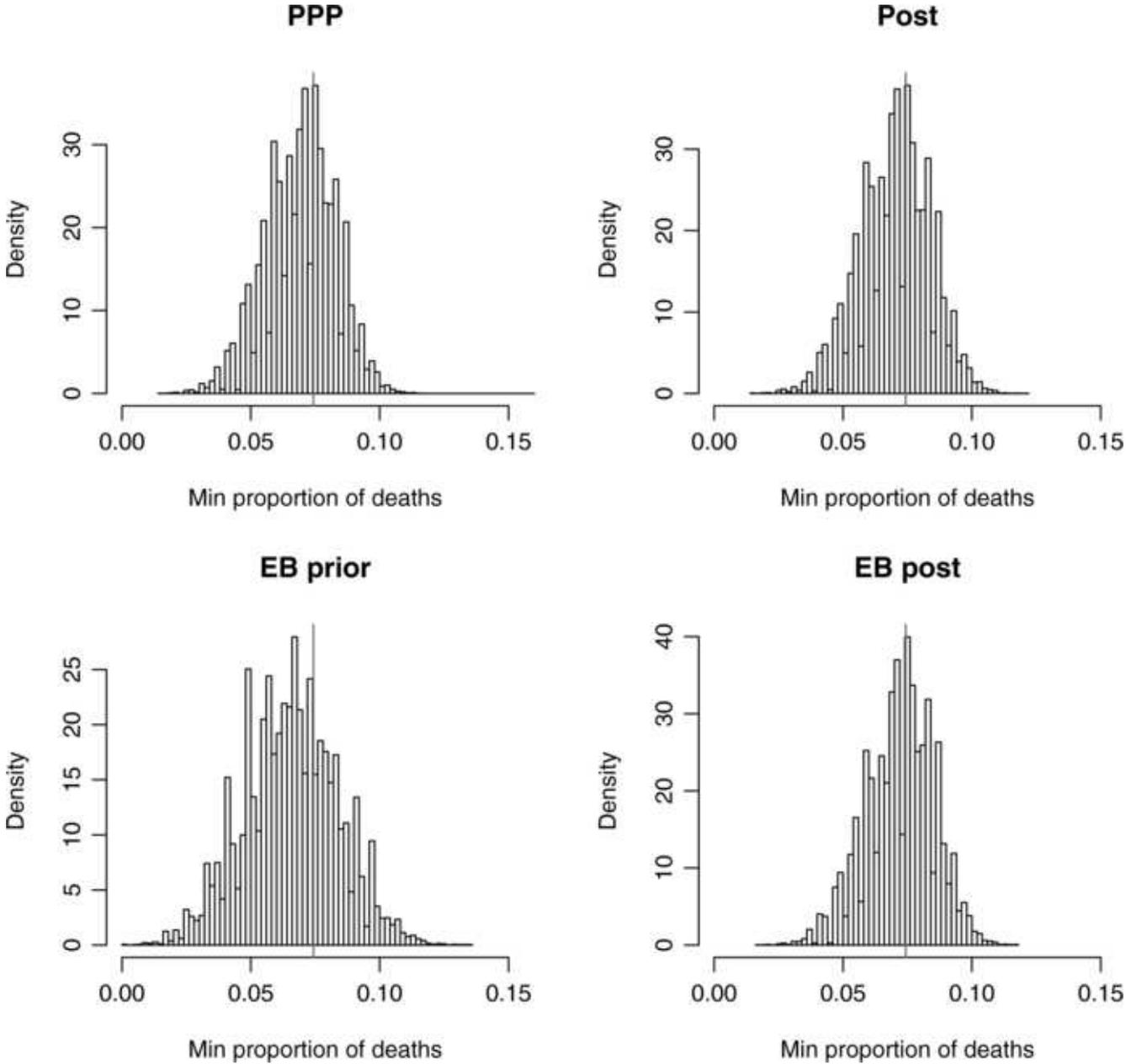

Fig. 6. *Predictive distribution for $T = \min\{y_i/n_i\}$ in the Bristol Royal Infirmary data.*

(C.4)
$$\propto \frac{\pi_{post}(\theta_i \mid \mu, \tau^2, \mathbf{x}_{obs})}{f(t_{obs} \mid \boldsymbol{\theta})}$$
$$\propto \frac{N(\theta_i \mid E_i, V_i)}{f(t_{obs} \mid \boldsymbol{\theta})},$$

where $E_i, V_i$ are given in (B.3). Next we need to find a good proposal to simulate from (C.4). An obvious proposal would simply be the posterior $\pi_{post}(\theta_i \mid \mu, \tau^2, \mathbf{x}_{obs})$, but this can be a very bad proposal when the data are indeed "surprising" for the entertained model. In particular, the posterior distri-bution centers around the MLE $\widehat{\boldsymbol{\theta}}$ while the partial posterior centers around the *conditional* MLE, $\widehat{\boldsymbol{\theta}}_{cMLE}$, that is,

$$\widehat{\boldsymbol{\theta}}_{cMLE} = \arg\max f(\mathbf{x}_{obs} \mid t_{obs}, \boldsymbol{\theta})$$
$$= \arg\max \frac{f(\mathbf{x}_{obs} \mid \boldsymbol{\theta})}{f(t_{obs} \mid \boldsymbol{\theta})}.$$

It is intuitively obvious that, when the data are not "surprising," that is, when $t_{obs}$ comes from the "null" model, then $f(\mathbf{x}_{obs} \mid t_{obs}, \boldsymbol{\theta})$ would be similar to $f(\mathbf{x}_{obs} \mid \boldsymbol{\theta})$ and the partial and posterior distributions would also be similar. However, when the data are "surprising" and $t_{obs}$ is not a "typical" value, then the "null" model and the conditional model can be considerably different, as well as the corresponding MLEs. For Metropolis proposals, Bayarri and Berger (2000) then suggest generating from the posterior distribution but then "moving" the generated values closer to the mode of the target distribution (the partial posterior) by adding

$$\widehat{\boldsymbol{\theta}}_{cMLE,i} - \widehat{\boldsymbol{\theta}}_{MLE,i},$$

multiplied (when this results in improved mixing) by a Uniform(0,1) random generation. This and other algorithms for computing conditional distributions are presented in Bayarri, Castellanos and Morales (2006).

To avoid computation of $\widehat{\boldsymbol{\theta}}_{cMLE}$, which can be rather time consuming, we use instead an estimate $\widetilde{\boldsymbol{\theta}}_c$ which we expect to be close enough (for our purposes) to $\widehat{\boldsymbol{\theta}}_{cMLE}$ for this model and this $T$ (see Bayarri and Morales (2003)). In particular, we take all components to be equal and given by

$$\widetilde{\theta}_c = \frac{\sum_{l=1}^{I-1} \overline{X}_{(l\cdot)}}{I-1},$$

where $(\overline{X}_{(1\cdot)}, \ldots, \overline{X}_{(I\cdot)})$ denote the group means sorted in ascendent order. That is, we simply remove the largest sample mean and then average (we could have also used a weighted average if the sample sizes were very different).

Then, the resulting algorithm to simulate from (C.4) at stage $k$, given the (simulated) values $(\boldsymbol{\theta}_{-i}^k, \theta_i^k, \mu^k, \tau^{2(k)})$, is:

1. Simulate $\theta_i^* \sim N(\theta_i \mid E_i, V_i)$.
2. Move the simulation $\theta_i^*$ to

$$\widetilde{\theta}_i^* = \theta_i^* + U \cdot (\widetilde{\theta}_c - \widetilde{\theta}_{MLE,i}),$$

where $U$ is random number in $(0,1)$.
3. Accept candidate $\widetilde{\theta}_i^*$ with probability

$$\alpha = \min\left\{1, \frac{N(\widetilde{\theta}_i^* \mid E_i, V_i)N(\theta_i^k \mid E_i, V_i)f(t_{obs} \mid \boldsymbol{\theta}_{-i}^k, \widetilde{\theta}_i^k)}{N(\widetilde{\theta}_i^k \mid E_i, V_i)N(\theta_i^* \mid E_i, V_i)f(t_{obs} \mid \boldsymbol{\theta}_{-i}^k, \widetilde{\theta}_i^*)}\right\}.$$

## APPENDIX D: DERIVATION OF THE FULL CONDITIONAL OF $\theta$'S IN SECTION 4.3

The full conditional partial posterior density for $\theta_i$ is

$$\pi(\theta_i \mid \tau^2, \theta_1, \ldots, \theta_{i-1}, \theta_{i+1}, \ldots, \theta_I, \mathbf{x}_{obs} \setminus t_{obs})$$

$$\propto \frac{\pi_{post}(\theta_i \mid \tau^2, \theta_1, \ldots, \theta_{i-1}, \theta_{i+1}, \ldots, \theta_I, \mathbf{x}_{obs})}{f(t_{obs} \mid \theta_1, \ldots, \theta_i, \ldots, \theta_I)}$$

$$\propto \exp\left\{-\frac{1}{2}\left(\frac{n_i}{\sigma_i^2} + \frac{1}{\tau^2}\right)\left(\theta_i - \frac{n_i\overline{x}_i\cdot/\sigma_i^2 + \mu_0/\tau^2}{n_i/\sigma_i^2 + 1/\tau^2}\right)^2\right\}$$

$$\cdot \exp\left\{\frac{1}{2}\frac{(\sum_j n_j)}{\sum_j n_j\sigma_j^2}\left(t_{obs} - \frac{\sum_{j=1}^I n_j\theta_j}{\sum_j n_j}\right)^2\right\}$$

$$\propto \exp\left\{-\frac{1}{2}\left(\theta_i^2\left(\frac{n_i}{\sigma_i^2} + \frac{1}{\tau^2}\right) - 2\theta_i\left(\frac{n_i}{\sigma_i^2}\overline{x}_i\cdot + \frac{1}{\tau^2}\mu_0\right)\right)\right\}$$

$$\cdot \exp\left\{\frac{1}{2\sum_j n_j\sigma_j^2}\left(\sum_j n_jt_{obs} - n_i\theta_i - \sum_{l \neq i} n_l\theta_l\right)^2\right\}$$

$$\propto \exp\left\{-\frac{1}{2}\theta_i^2\left(\left(\frac{n_i}{\sigma_i^2} + \frac{1}{\tau^2}\right) - \frac{n_i^2}{\sum_j n_j\sigma_j^2}\right)\right.$$

$$- 2\theta_i\left(\frac{n_i}{\sigma_i^2}\overline{x}_i\cdot + \frac{\mu_0}{\tau^2}\right.$$

$$- \frac{n_i}{\sum_j n_j\sigma_j^2}$$

$$\left.\left.\cdot\left(\sum_j n_jt_{obs} - \sum_{l \neq i} n_l\theta_l\right)\right)\right\},$$

which, after some algebra, reduces to

$$\pi(\theta_i \mid \tau^2, \theta_1, \ldots, \theta_{i-1}, \theta_{i+1}, \ldots, \theta_I, \mathbf{x}_{obs} \setminus t_{obs})$$

$$\propto \exp\left\{-\frac{1}{2V_i^0}(\theta_i - E_i^0)^2\right\},$$

with $E_i^0$ and $V_i^0$ given in (4.6) and (4.7), respectively. The result then follows if $V_i^0$ can be shown to be greater than 0, which is true because $1 - \frac{n_i\sigma_i^2}{\sum_{j=1}^I n_j\sigma_j^2} > 0$.

## ACKNOWLEDGMENTS

This work was supported by the Spanish Ministry of Science and Technology Grants MTM2004-03290 and TSI2004-06801-C04-01.